\documentstyle[preprint,aps,eqsecnum]{revtex}

\begin{document}
\draft

\title{Delays Associated with Elementary Processes \\
in Nuclear
Reaction Simulations}
\author{Pawe\l \ Danielewicz$^{1,2}$ and Scott Pratt$^1$}
\address{$^1$National Superconducting Cyclotron Laboratory and
Department of Physics and Astronomy\\
Michigan State University, East Lansing MI  48824}
\address{$^2$Grand Acc\'el\'erateur National d'Ions Lourds
(GANIL), BP 5027\\
F-14021 Caen Cedex, France}
\maketitle

\begin{abstract}
Scatterings, particularly those involving resonances, and other elementary
processes do not happen instantaneously.  In~the context of semiclassical
nuclear reaction simulations, we consider delays associated with an
interaction for incident quantum wave-packets.  As~a consequence, we express
delays associated with elementary processes in terms of elements of the
scattering matrix and phase shifts for elastic scattering.  We~show that, to
within the second order in density, the simulation must account
for delays in scattering consistently with the mean field in
order to properly model thermodynamic properties such as
pressure and free-energy density.  The~delays associated with
nucleon-nucleon and pion-nucleon scattering in free space are
analysed with their nontrivial energy dependence.  Finally,
an~example of $s$-channel scattering of massless partons is
studied, and scattering schemes in nuclear reaction simulations
are investigated in the context of scattering delays.
\end{abstract}


%%%%%%%%%%%%%%%%%%%%%%%%%%%%%%%%%%%%%%%%%%%%%%%%%%%%%%%%%%%%%%%
\narrowtext

\section{Introduction}

Semiclassical transport-simulations have provided the theoretical backbone for
interpreting heavy-ion reactions at energies in excess of few tens of MeV per
nucleon.  Simulations follow either a~cascade
approach\cite{yar79,cug80} or a~solution to the Boltzmann
equation including the mean
field and the Pauli principle\cite{won82,ber84,ber88}.  Within these
approaches particles move classically in-between collisions where momenta
change abruptly and possibly new particles form.  In~practice, the Pauli
principle appeared primarily important for reactions below
150~MeV/nucleon.
For beam energies of hundreds of MeV per nucleon, detailed comparisons of
simulations with data have been carried out\cite{gal87,aic87,pan93,zha94} in
order to determine features of the momentum and density dependence of the
nuclear mean field.  Within many-body theory, efforts were made to calculate
nuclear optical potentials starting from elementary NN
interactions\cite{ter87,wir88,fri81}.  In~describing reactions
up to the beam energies
$\sim$2~GeV/nucleon, pions and low-lying resonances have also
been included in simulations.

It has been noted that results of heavy-ion reaction simulations can depend
sensitively on prescriptions of space-time details of elementary scattering
processes\cite{hal81,gyu82,mal84,sch91,zha94,kah94}.  Such details of
elementary processes may matter even more at higher beam energies than at those
of the MSU or SIS accelerators, specifically, at those of AGS
(14~GeV/nucleon),
SPS (200~GeV/nucleon), and of the constructed and planned
colliders, RHIC and
LHC. This is because of high Lorentz dilation factors that can amplify
space-time effects, and because of high particle densities early in the
reactions.

In this paper we investigate the time duration of elementary processes in free
space, in~the context of semiclassical simulations.  Both the
forward-going and
scattered waves get delayed in elementary processes.  While delay of the
forward wave is consistent with effects of an~optical potential on a~particle
in a~medium, to~the lowest order in particle density, both
types of delay
times, of the forward and scattered waves, affect the density of states in
energy, to~the lowest nontrivial second-order in particle
density.  As~one
important outcome of our work, we find that, to describe the density of states
and associated thermodynamic quantities such as pressure, heat capacity, etc.,
the delays for scattered particles must be properly accounted for in
semiclassical simulations, in~addition to the effects of the optical potential.
If one relaxes the requirements on interactions in simulations,
forsaking e.g.~the proper behavior of single-particle energies,
but demanding that free energy
as a~fundamental thermodynamic function, is properly reproduced to within the
second order in density in equilibrium, then one can manipulate the mean field
to entirely absorb effects of the delays for scattered particles.
Alternatively, time delays for scattered particles may be manipulated to absorb
the effects of optical potential.  Even in a~degenerate Fermi gas, as we show,
the delays in scattering can affect thermodynamic quantities.  We~investigate
several examples of particle-particle scattering amplitudes describing,
respectively, the scattering through a~sharp Breit-Wigner
resonance, the~$\pi
N$ interaction in the $\Delta$ channel, and the $s$-channel scattering of
massless partons.  Using free-space amplitudes, we calculate numerically the
time delays for the $NN$ system.  We~show that the delay for scattering has
some dependence on the scattering angle.  The~delays become small at high
energies.  Finally, we investigate scattering schemes in
simulations\cite{cug80,ber84,ber88,hal81,gyu82,mal84,sch91,kah94}
in the context of time
delays.  The~schemes include hard-sphere scattering and scattering at the
distance of closest approach.  We~outline numerical strategies for determining
the equations of state and transport coefficients to an~arbitrary order in
density given the scattering prescription.

The content of subsequent sections is as follows.
In~Sec.~\ref{Time} we
consider a~wave packet traversing a~region with a~scatterer and derive
expressions for the delays of scattered and forward-going waves, in~terms of
scattering amplitudes (phase shifts) and their derivatives.
In~Sec.~\ref{Ergo}
we demonstrate that these delay times are consistent with an~ergodic constraint
stating that the extra time spent in the vicinity of a~scatterer should be
proportional to the change in the density of states brought about by that
scatterer.  We~further show that the delay time for a~forward-going wave
corresponds to the classical motion of a~particle in a~mean field given by the
self-energy for a~plane wave in a~low-density medium.  We~also
discuss in that section and more in the Appendix, in~the
context of ergodic constraints, the~modification of the
scattering of two
bodies brought about by the presence of a~third body.  Finally,
we~express
the free-energy density and pressure in terms of different time
delays.  In~Sec.~\ref{Stat} we draw conclusions on the time
delays in a~degenerate medium
by examining an~expression for the change in the pressure due to two-particle
interactions.   Section~\ref{Res} is devoted to the examples of
resonance
scattering.  In~ Sec.~\ref{NNScat} we calculate spin-isospin averaged time
delays for $NN$ system and directly relate the delays to the second virial
coefficient.  Different scattering prescriptions and methods for the
determination of equations of state and transport coefficients,
to~within
arbitrary order in density for given prescriptions, are
discussed in~Sec.~\ref{Simu}.

\section{Time Delays from Wave-Packet Dynamics}
\label{Time}

We consider a~wave packet incident on a~large spherical volume of radius $R$,
at the center of which interaction takes place.  To~make the
derivation simpler
we assume that this volume is much larger than the wave packet.  The~wave
function is
\begin{equation}
\psi_{\alpha} (t) = \int dE\ g(E)\ \psi_{\alpha}^E (t),
\end{equation}
where for large distances
\begin{eqnarray}
\nonumber
\psi_{\alpha}^E & = & \sum_{\beta} \phi_{\alpha \beta} ({\bf
r},t) \, \chi_{\beta}\\
& = & \phi_{\alpha}^0 ({\bf r},t) \, \chi_{\alpha} +
\sum_{\beta} f_{\alpha
\beta} (\theta, E) \, {{\rm e}^{i(k_{\beta} r - Et)} \over r}
\, \chi_{\beta} ,
\end{eqnarray}
and
\begin{equation}
\phi_{\alpha}^0 = {1 \over 2i k_{\alpha} r}
\sum_{\ell = 0}^{L_{max}} (2\ell+1) P_{\ell} (\cos{\theta})
\left[ (-1)^{(\ell +1)}
{\rm e}^{-i(k_{\alpha} r + Et)} + {\rm e}^{i(k_{\alpha} r -
Et)} \right] .
\label{inci}
\end{equation}
Here $\alpha$ and $\beta$ denote the initial and final channel,
respectively, and $\chi$ is the internal wave function.  Most
often we shall consider elastic scattering of particles with no
internal degrees of freedom, hence
\begin{equation}
f(\theta, E) = - {1\over 2k} \sum_{\ell} \, (2\ell + 1) \,
P_{\ell} (\cos{\theta}) \, {\cal T}_{\ell} (E),
\label{fteta}
\end{equation}
where the amplitude ${\cal T} _{\ell}$ is related to scattering matrix
and phase shift by
\begin{equation}
{\cal T} _{\ell} = i (S_{\ell} - 1) = -2 \sin{\delta_{\ell}}
\exp i \delta_{\ell} , \label{phase}
\end{equation}
$S_{\ell} = \exp
2i\delta_{\ell}$.  For~the asymptotic form to be valid $r$ must
be much larger than $L_{max}/k$ and $T_{\ell}$ must vanish
for $\ell > L_{max}$.

The average time $\tau_{vol}$ that
a particle spends in the volume can be written as
\begin{equation}
\tau_{vol} = R^2 \, {\sum_{\gamma} \int dt \, d\Omega \, {\bf
j}_{\gamma} (R, \Omega, t) \cdot \hat{\bf r} \, t \over N} ,
\label{avtim}
\end{equation}
where current density in channel $\gamma$ is
\begin{equation}
{\bf j}_{\gamma} = \frac{1}{2
\mu_{\gamma} i} \, \phi_{\gamma}^{*}({\bf r},t) \,
{\buildrel \leftrightarrow \over \partial} \, \phi_{\gamma}
({\bf r},t) ,
\end{equation}
and $N$ is integrated incident flux through the surface.
The~time in~(\ref{avtim})
is represented as the average time for the particle
to leave the
volume minus the average time for the particle to enter the
volume, as ${\bf j} \cdot \hat{\bf r}$ is positive when
a particle exits and negative when a~particle enters.
We decompose ${\bf j}$ into incoming and outgoing pieces and,
as there is no interference between incoming and outgoing waves
for a~large volume, we write explicitly
\begin{equation}
\tau_{vol} = \tau_{out} - \tau_{in}.
\end{equation}
With the incoming wave having only a~contribution from the
unscattered
wave packet, one finds, from (\ref{inci}) and (\ref{avtim}),
that $\tau_{in} = -R/v_{\alpha}$.

The outgoing wave has contributions from both the scattered and
unscattered portions of the wave packet.  Correspondingly,
the outgoing current has three contributions:
from the scattered wave, from the
unscattered wave, and from the interference between the two.
We~first calculate the exit time associated with the scattered
wave in the single-channel case, for particles leaving at the
center of mass angle $\theta$, using partial-wave expansion,
\begin{eqnarray}
\nonumber
\tau_{s}(\theta) &=& R^2 \, \frac{ \int dt \,
{\bf j_{s}}(R,\theta,t) \cdot {\hat{\bf r}} \,
t}{dN_s/d\Omega}\\ \nonumber
&=& \frac{1}{Z} \sum_{\ell, \ell^{\prime}}^{L_{max}}
\int dt\, dE \, dE^{\prime} \, g(E) \, g(E^{\prime})\,
{\rm e}^{i(k-k^{\prime})R}
\, (2\ell+1) \, P_{\ell}(\cos{\theta})
\\
&&\times (2\ell^{\prime}+1) \, P_{\ell^{\prime}}(\cos{\theta})
\, {\cal T}^{*}_{\ell^{\prime}}(E^{\prime}) \, {\cal
T}_{\ell}(E) \,  t \, {\rm e}^{-i(E-E^{\prime})t}
\left(\frac{1}{k} + \frac{1}{k^{\prime}}\right),
\end {eqnarray}
where $Z$ is the same expression as the sum on the r.h.s.~of
the last equation, only without the time $t$ in the integrand,
\begin{eqnarray}
\nonumber
Z &=& \sum_{\ell,\ell^{\prime}}^{L_{max}}
\int dt\,  dE \, dE^{\prime} \, g(E) \, g(E^{\prime}) \,
{\rm e}^{i(k-k^{\prime})R} \,
(2\ell+1) \, P_{\ell} (\cos{\theta})
\\
&&\times
(2\ell^{\prime}+1) \, P_{\ell^{\prime}}(\cos{\theta}) \,
{\cal T}^{*}_{\ell^{\prime}}(E^{\prime}) \,
{\cal T}_{\ell}(E) \, {\rm e}^{-i(E-E^{\prime})t}
\left(\frac{1}{k} + \frac{1}{k^{\prime}}\right).
\end{eqnarray}
By making the substitution,
\begin{equation}
t=\frac{i}{2}\left\{ \frac{\partial}{\partial E} - \frac{\partial}{\partial
E^{\prime}} \right\},
\end{equation}
then carrying out integration by parts to make the derivatives
with respect to  $E$ and
$E^{\prime}$ act~on~$g$, $\exp{i(k-k^{\prime})R}$, and~${\cal
T}$, and further
integrating over the time coordinate and one of the energy variables, one
obtains a~simple expression for~$\tau_{s}$.  For~a
well-defined energy this expression is
\begin{eqnarray}
\nonumber
\tau_{s}(\theta ) &=& {R \over v} + \Delta \tau_s (\theta ) \\
\nonumber
& = &
\frac{R}{v} + \frac{1}{Z^{\prime}}
\sum_{\ell,\ell^{\prime}}
^{L_{max}}(2\ell+1) \, P_{\ell}(\cos{\theta})\,
(2\ell^{\prime}+1) \, P_{\ell^{\prime}}(\cos{\theta})
\\
&&\times \frac{1}{2i}\left[
{\cal T}^{*}_{\ell^{\prime}}(E) \left( \frac{d}
{d E} {\cal T}_{\ell}(E)\right) -  \left(
\frac{d}
{d E} {\cal T}^{*}_{\ell^{\prime}}(E)\right)
{\cal T}_{\ell}(E)\right] ,
\label{taust}
\end{eqnarray}
where
\begin{equation}
Z^{\prime} = \sum_{\ell,\ell^{\prime}}^{L_{max}}
(2\ell+1) \, P_{\ell} (\cos{\theta})\
(2\ell^{\prime}+1) \, P_{\ell^{\prime}}(\cos{\theta}) \,
{\cal T}^{*}_{\ell^{\prime}}(E) \, {\cal T}_{\ell} (E) .
\end{equation}
The more general result when many channels are open, in~terms
of the scattering amplitude~$f$,~is
\begin{equation}
\tau_s^{\alpha \beta} (\Omega ) = {R \over v_{\beta}} + {1
\over 2 i |f_{\alpha \beta} |^2 } \left\{ f_{\alpha \beta}^*
\, {d f_{\alpha \beta} \over d E} -
f_{\alpha \beta} \,
{d f_{\alpha \beta}^* \over d E} \right\} .
\label{tausa}
\end{equation}
This is of use when determining time delays for a~NN system in
 Sec.~\ref{NNScat}.  In~that case the indices refer to spin
components.

For the case of elastic scattering in only one partial wave,
the exit time from (\ref{taust}) may be written as
\begin{equation}
\tau_{s} = \frac{R}{v} + \frac{d\delta_{\ell}}{dE}.
\label{taus}
\end{equation}
Thus, the extra time delay due to the scattering is $d\delta
/dE$ which is not
the naive guess that one would make from considering an~incoming partial wave
reflecting off a~potential.  In~that case the answer should be
$2 \, d\delta /dE$
since the potential modifies the scattered wave by
a~factor~${\rm e}^{2i\delta}$.
This  discrepancy is associated with the interference between
the forward-going wave and the scattered wave.

Next, we consider the current and the average exit time
associated with
the interference of the forward wave $\phi^0$ with itself and
with the scattered wave.  The~angular integration must extend
up to
$\theta \simeq L_{max}/kR$ which limits the forward wave,
and it may, in~particular, extend over the whole angular range.
Analogous procedure to that before yields
\begin{equation}
\tau_{f} = \frac{R}{v} - \frac{1}{2Z^{\prime\prime}}
\sum_{\ell=0}^{L_{max}} (2\ell+1)
 \frac{d}
{d E} \left( {\cal T}_{\ell} (E) +
 {\cal T}^{*}_{\ell} (E) \right)
\label{tauf}
\end{equation}
where
\begin{equation}
Z^{\prime\prime} = \sum_{\ell=0}^{L_{max}} (2\ell+1)
\left(1-i({\cal T}_{\ell}
(E)-{\cal T}_{\ell}^{*}(E))\right) \end{equation}
In terms of phase shifts this gives
\begin{equation}
\tau_{f} = \frac{R}{v} + \frac{\sum_{\ell}^{L_{max}} \,
(2\ell+1) \, 2 \cos{2\delta_{\ell}} \,
\frac{d\delta_{\ell}}{dE}}{\sum_{\ell}^{L_{max}} \,
(2\ell+1) \, (1-4\sin ^2 {\delta_{\ell}})}.
\label{forwanswer}
\end{equation}
The more general result in terms of the scattering
amplitude~$f$~is
\begin{equation}
\tau^{\alpha}_f = {R \over v_{\alpha}} + {2
\over
\sum_{\ell}^{L_{max}} (2 \ell + 1) - 4 k_{\alpha} \, {\rm Im}
\, f_{\alpha \alpha} (0)} \,
{d \over d E}
\Big\lbrace k_{\alpha} \, {\rm Re} \, f_{\alpha \alpha} (0)
\Big\rbrace
{}.
\label{taufa}
\end{equation}

By combining the results for scattered and forward waves,
we~obtain the average exit time in the single-channel case equal
to
\begin{eqnarray}
\nonumber
\tau_{out} &=& \tau_{f}\left(1-\frac{N_{s}}{N}\right)
 + \int d\Omega \, \tau_{s}({\Omega}) \, \frac{dN_s/d\Omega
}{N}\\
&=& \frac{R}{v} + \frac{\sum_{\ell}^{L_{max}} \, (2\ell+1)
\, 2 \, \frac{d\delta_{\ell}}{dE}}{\sum_{\ell}^{L_{max}} \,
(2\ell+1)}.
\label{taut}
\end{eqnarray}
The total number of scattered particles is $N_s$.  The~delay
time~(\ref{taut})
is the same that one would have guessed assuming that different partial waves
acted independently and each was delayed by~$2d\delta_{\ell}
/dE$.  The~complexity of Eqs.~(\ref{forwanswer}) and (\ref {taust}) stems from
interference of partial waves.  In~the next section we discuss the consistency
of these results with equilibrium expectations.  We~show that the forward delay
time agrees with a~delay in the motion through a~mean field.

\section{The Ergodic Constraint and the Mean Field}
\label{Ergo}

\subsection{Ergodicity}
\label{Erg}

All thermodynamic variables such as the pressure can be found if one knows the
density of states within a~system.  Assuming that particles do not interact,
and that statistical effects can be ignored, one can obtain thermodynamic
quantities to lowest order in particle density.  Assuming that particles
interact only two at a~time allows one to calculate these quantities to the
next order in the density.  To account for the interaction of the two particles
at a~time, one needs to find the correction to the density of states of
relative motion.  This correction to the density of states
$\rho(E) \equiv d{\sl n}/dE$ is given
by phase shifts for two-particle scattering\cite{lan58,mek78},
\begin{equation}
\rho(E) = \rho_0 (E) + \Delta \rho (E) = \frac{4\pi V
}{(2\pi )^3} { k^2 \over v}
+ \frac{1}{\pi}\sum_{\ell} \, (2\ell +1) \, \frac{d\delta
_{\ell}}{dE} ,
\label{rhoe}
\end{equation}
where $k$ is the reduced relative momentum, $V$ is the volume, and $v$ is the
velocity of relative motion.  This is a~complete quantum-mechanical answer in
the single-channel case.  If~processes such as $a+b
\rightarrow c+d$ can occur, a~similar answer can be obtained by
diagonalizing the S-matrix and finding the eigenphases\cite{pra87}.

In a~simulation, the equations of motion should result in a~change in the
probability for two particles of a~specific energy to be in the vicinity of
each other, proportional to the change in the density of states for these
particles.  If~the equations of motion accomplish this, the classical
simulation will yield correct thermodynamic quantities.
We~consider a~subvolume~$V$ in relative-coordinate space and
a~narrow relative-energy range
$|E - E_0| < \delta E$.  Given that particles spend a~portion of time $\tau_0$
within $V$ in the absence of interactions, when sampling is carried out over a
long time $T$, then, in~the presence of interactions, the time within $V$
should change by~$\Delta \tau$ such that
\begin{equation}
\frac{\Delta \tau}{\tau_0} =  {\Delta \rho \over \rho_0} =
\frac{\frac{1}{\pi}\sum_{\ell} \, (2\ell+1) \,
\frac{d\delta _{\ell}}{dE}}{\rho_0} .
\label{delt}
\end{equation}
This can be thought of as an~ergodic constraint, since a~system continuously
sampled over a~long time should be populated according to its contribution to
the density of states.  In~this section we explore the case of elastic
scattering and wish to see if the results of the preceeding section are
consistent with the ergodic constraint and with the picture of particles moving
through a~mean field.

In the context of a~classical simulation, the additional time $\Delta \tau$
will come from three causes.  First, particle kinetic energy within the volume
$V$ will be different as compared to free space, by the negative of the mean
potential within the volume, $-u(k)$.  The~change in the energy gives a~change
in the momentum and in the velocity and, in~consequence, in~the time spent
within subvolume by an~amount denoted $\Delta \tau _{1}^{clas}$.  In~addition,
the energy dependence of the mean potential generally changes the velocity by
$\Delta v = du/dk$.  The~corresponding change in the time spent within the
volume is denoted as $\Delta \tau_{2}^{clas}$.  Both these contributions to the
time follow from classical equations of motion within a~potential.  A~third
contribution $\Delta \tau^{coll}$ stems from collisions.  Because of collisions
the particles will emerge on the average earlier or later (particularly in the
case of resonance scattering) from the volume than in absence of collisions.
Summing the three contributions should yield
\begin{equation}
\frac{\Delta \tau_{1}^{clas} + \Delta \tau_{2}^{clas} + \Delta
\tau^{coll}}{\tau_0} =
\frac{\frac{1}{\pi}\sum_{\ell} \, (2\ell +1) \, \frac{d\delta
_{\ell}}{dE}}{\rho_0}.
\label{ergcl}
\end{equation}

The time $\Delta \tau_{1}^{clas}$ can be easily obtained from the ergodic
theorem for a~{\em classical} potential.  Otherwise, one can resort to
geometric considerations caring, however, for the fact that, to~the lowest
order in $u$, the velocity changes its direction as well as the magnitude.  As
the relative change in time spent within the subvolume should be proportional
to the relative change in the density of states brought about by the change in
kinetic energy, we get
\begin{equation}
\frac{\Delta \tau_{1}^{clas}}{\tau_0} = \frac{\Delta
\rho_0}{\rho_0}
= - \frac{u(E) \, \frac{d\rho_0}{dE}}{\rho_0} .
\label{t1def}
\end{equation}
The contribution $\Delta \tau_{2}^{clas}$ is obtained by noting
that the time it
takes to traverse a~given path through the subvolume
is proportional to the inverse of velocity,
\begin{equation}
\frac{\Delta \tau_{2}^{clas}}{\tau_0} = -\frac{\Delta v}{v}
= -{1 \over v} \, \frac{du}{dk}
= -\frac{du}{dE}.
\label{t2def}
\end{equation}
The contribution due to collisions, $\Delta \tau^{coll}$, will
be determined by
the collision rate and the change~$\Delta \tau_s$ in the
average time spent
within the collision range compared to free flight time,
\begin{equation}
\Delta \tau^{coll} = \tau_0 \, {\sigma v \over V} \,
\Delta \tau_s .
\label{tauc}
\end{equation}
Combining the expressions for different contributions to the change in time on
account of interactions, we~find that, according to the ergodic
constraint, we~should have
\begin{equation}
\frac{1}{\pi}\sum_{\ell}(2\ell+1)\frac{d\delta_{\ell}}{dE} =
-\frac{d\rho_0}{dE} \, u
-\rho_0 \, \frac{du}{dE}
+ \rho_0  \, { \sigma v \over V} \, \Delta \tau_{s}.
\label{erg2}
\end{equation}

In the following we shall express~$u$ in terms of phase shifts
and then
determine~$\Delta \tau_{s}$ from~(\ref{erg2}) and examine
whether the result is
consistent with what was obtained in preceeding section.
The~potential $u$ in~(\ref{erg2}) is identified with the
correction to the kinetic
energy at which the single-particle Green's function has a~pole,
\begin{eqnarray}
\nonumber
g(k_1,E_1) & = &  \frac{1}{E_1 - e(k_1) - u + i\gamma/2} \\
& = & \frac{1}{
E_1 - e(k_1) - \langle{\bf k}|{\cal T} |{\bf k}\rangle
/ V} ,
\label{gf}
\end{eqnarray}
where we use the fact that self-energy can be expressed in terms of the forward
element of the ${\cal T}$-matrix and where
$k$ continues to be the relative particle
momentum.  Assuming that angular momentum is conserved,
we~obtain
\begin{eqnarray}
\nonumber
{1 \over V} \, \langle {\bf k}|{\cal T}|{\bf k} \rangle
& = & {1 \over V} \, \sum_{\ell ,m} \,
\langle {\bf k}|\ell ,m \rangle \langle \ell ,m|{\cal T}
 |\ell ,m \rangle \langle \ell ,m|{\bf k} \rangle \\
& = & \sum_{\ell} \, (2\ell +1) \, \frac{{\cal T}_{\ell}
}{2\pi \rho_0} .
\end{eqnarray}
The imaginary part of the matrix times~$-2$ can be shown to be
equal to $\sigma v
$ by using the standard expression for the cross section involving
$\sin^2 {\delta}$.  Then $\gamma$ is indeed the geometric
scattering rate one expects.
Turning now to the real part of ${\cal T}$-matrix, the potential associated
with the presence of other particle within subvolume becomes
\begin{equation}
u(E) = - \, \frac{ \sum_{\ell} \, (2\ell + 1)
\sin{2\delta_{\ell}}}{2 \pi \rho_0} .
\label{udef}
\end{equation}

The sum of the times $\Delta\tau_{1}^{clas}$ and
$\Delta\tau_{2}^{clas}$ in~(\ref{erg2}) involves the derivative
$d\left(u(E)\rho_0 (E)\right) /dE$.  With
(\ref{udef}), we obtain
\begin{equation}
\frac{\Delta\tau_{1}^{clas} + \Delta\tau_{2}^{clas}}{\tau_0}
= - \, \frac{1}{\rho_0} \, \frac{d}{dE} \left(\rho_0 \,
u(E)\right) = \frac{1}{\pi\rho_0} \sum_{\ell} \,
(2 \ell + 1) \, \cos{2\delta_{\ell}}
\,\frac{d\delta_{\ell}}{dE} .
\label{clasum}
\end{equation}
Upon inserting the above result into Eq.~(\ref{erg2}) we find
\begin{equation}
\Delta \tau_{s} =
{
\sum_{\ell} \,
(2\ell+1) \, \sin^2{\delta_{\ell}} \, \frac{d\delta_{\ell}}{dE}
\over
 \sum_{\ell} \, (2\ell+1) \, \sin^2{\delta_{\ell}}
}
{}.
\label{tauda}
\end{equation}
If only one partial wave is scattered, then the change in time reduces to
\begin{equation}
\Delta \tau_{s} = \frac{d\delta_{\ell}}{dE}.
\end{equation}
This result is consistent with Eq.~(\ref{taus}) of the last
section. If~many
waves are scattered, then there is generally interference between various
scattered partial waves giving rise to a~variation of the change in time with
scattering angle, cf.~Eq.~(\ref{taust}).
Equation~(\ref{tauda}) represents, in~such a~case, an~average
of the change of time over the scattering angles,
weighted with the flux or cross section.

One can further see that the time delay of the forward wave, as derived in the
last section, may be identified with $\Delta \tau_{1}^{clas} +
\Delta \tau_{2}^{clas}$.  The wave packet
in the last section was of a~transverse
size~$B=L_{max}/k$.  We~further assume that the length of the
packet is~$W$.
The~time for such a~packet to move by the potential in the
absence of
interactions is~$\tau_0 = W/v$.  Using Eq.~(\ref{tauf}) in the
limit of large~$L_{max}$ we obtain
\begin{equation}
\frac{\Delta \tau_{f}}{\tau_0} = \frac{\sum_{\ell} \, (2\ell+1)
\, 2 \cos{2\delta_{\ell}} \,
\frac{d\delta_{\ell}}{dE}}{\frac{W}{v}\, \sum_{\ell} \,
(2 \ell+1)}.
\label{deltaf}
\end{equation}
For wide packets we can approximate $\sum_{\ell} \, (2\ell+1)
\simeq 2\int \ell \, d\ell
= k^2 \, B^2$ and, with the volume of a~packet being equal
to~$V = \pi B^2 W$, we~can rewrite~(\ref{deltaf})~as
\begin{equation}
\frac{\Delta \tau_{f}}{\tau_0} = \frac{\sum_{\ell} \, (2\ell+1)
\, \cos{2\delta_{\ell}} \,
\frac{d\delta_{\ell}}{dE}}{\frac{W \, B^2 k^2}{v}}
= \frac{\sum_{\ell} \, (2\ell+1) \, \cos{2\delta_{\ell}} \,
\frac{d\delta_{\ell}}{dE}}{\pi \rho_0},
\label{foranswer}
\end{equation}
where $\rho_0$ is the density of states inside $V$.  One
can see that this agrees with (\ref{clasum}).

Given a~uniform many-body system, with density $n$ and temperature $T$ treated
as independent variables, all intensive thermodynamic quantitates can be
obtained from the free energy per unit volume ${\sl f}$.  In~terms of the
correction to the density of states of relative motion in energy in
(\ref{rhoe}), the free energy ${\sl f}$, to~the second order in the density,
can be written~as
\begin{equation}
{\sl f} (n,T) = {\sl f} _0(n,T) + \Delta P (n,T) ,
\label{fnT}
\end{equation}
where ${\sl f} _0$ is free energy for a~noninteracting system and $\Delta P$ is
the correction to pressure\cite{bet37,lan58}
\begin{equation}
P = P_0 + \Delta P = n \, T - T \, n^2  \, \left( {4 \pi \over
m T} \right)^{3/2} \, {1 \over 2} \int dE \, {\rm e}^{-E/T}\,
{1 \over
\pi} \sum_{\ell} \, (2\ell + 1) {d \delta_{\ell} \over dE} ,
\label{presde}
\end{equation}
and where the effects of statistics are ignored.  It can verified by a~direct
calculation that the correction to pressure may be further expressed as
\begin{eqnarray}
\Delta P & = & {1 \over 2}
\int {d {\bf p}_1 \over (2 \pi)^3}
\, { d {\bf p}_2  \over (2 \pi)^3}
\,  f (p_1) \,
f (p_2) \, {\rm Re} \left\langle ( {\bf p}_1 - {\bf p}_2
)/ 2  \right| {\cal T} \left| ( {\bf p}_1 - {\bf p}_2 )/ 2
\right\rangle \nonumber\\
& & - T \, {1 \over 2}
\int {d {\bf p}_1 \over (2 \pi)^3}
\, { d {\bf p}_2  \over (2 \pi)^3}
\,  f (p_1) \,
f (p_2)  \int d\Omega \, {d \sigma \over d \Omega} \, v \,
\Delta \tau_s (\Omega) .
\label{deltap}
\end{eqnarray}
Here $f = {\rm e}^{(\mu - E)/T} $ is phase-space occupancy.  The~first term on
the r.h.s.~of~(\ref{deltap}) accounts for the forward time delay or mean field.
The~second term accounts for delay in scattering and, depending
on the sign of~$\Delta \tau_s$, allows further for
an~interpretation in terms of the reduction
in the number of degrees of freedom or in terms of excluded volume.

\subsection{Limitation of the Considerations}

Two important conclusions can be reached regarding the method of the previous
section.  First, the sum of time delays for scattered and forward-going waves
is consistent with ergodic constraints.  Secondly, the time delay of the
forward-going wave is consistent with motion of a~particle through a~mean
field.  However, this consistency was derived assuming that the size of the
region used to determine the densities was large.  The~limitations of this
approximation are discussed below.

A problem with incorporating a~mean field which depends on the number of
particles within a~given volume element, is that such field can result in
classically bound states for attractive potentials, or inpenetrable potential
barriers for repulsive potentials.  Considerations in this section required
that the potential $u(E)$ was much smaller than any characteristic energy.
This is necessary for $\Delta \rho
\ll \rho_0$.
For a~sufficiently large volume $V$
this is not a~problem.  However, for a~finite $V$ there exists
phase space that is either bound
(attractive potentials) or unavailable (repulsive potentials),
of a~magnitude
\begin{equation}
\Delta = \frac {V}{(2\pi )^3} \frac{4\pi }{3}
\left( 2mu(k=0 ) \right)^{3/2}.
\end{equation}
On using $\sin{\delta_0} \approx - ka$, where $a$ is scattering
length, for low
$k$ in the expression for optical potential (\ref{udef}), one finds that
\begin{equation}
\Delta = \frac{4}{3\pi^{1/2}} \frac{a^{3/2}}{V^{1/2}} .
\end{equation}
Thus one finds that for a~{\em strict} validity of our considerations the
volume $V$ containing a~single scatterer must be chosen large compared to the
scattering length.  On examining Eq.~(\ref{tauf}) from the previous section,
one finds that, in~the first place, such a~condition must be satisfied to allow
for a~proper definition of $\tau_f$.  Scattering lengths are of the order of
the interaction range, unless a~resonance or bound state exist close to the
threshold.
Unfortunately, this is precisely the case for nucleon-nucleon
scattering near threshold where scattering lengths approach
20~fm.

\subsection{Scattering in the Presence of Third Bodies}
\label{Sati}

When considering the interaction of two particles in
Secs.~\ref{Time} and~\ref{Erg}, we~neglected the probability
that a~third body could interact with any
of the two particles within the subvolume used to define the mean field.
In~Eq.~(\ref{ergcl}) the contributions to the extra time spent
within the
subvolume, $\Delta \tau_1^{clas}$ and $\Delta \tau_2^{clas}$,
arose because the
particle trajectories were modified by the mean field.
The~first contribution
arose because the kinetic energy and direction
were altered
on~entering the mean field.  The~second contribution arose
due to dependence of
the mean field
on~momentum that led to the change of velocity by~$du(E)/dE$.
When many
scatterings occur, the~time spent in a~given relative state
does not just depend on~trajectories and velocity, but~further
on the manner in which such state is populated and depopulated.
If~population
of different states is to be consistent with first two of the
time delays in~Eq.~(\ref{ergcl}) due to interactions,
the~scattering prescriptions
need to be modified in the presence of third bodies.

Transition rates per unit time from
an~initial state, are given though, generally,
by~a~transition
matrix element to a~final state squared multiplied by the
final-state
density in energy.  For~example, in~the case of particles~1
and~3 scattering in the presence of particle~2, the~final-state
density would be obtained from a~product of single-particle
densities for~1 and~3.  Latter densities, up to a~factor, are
identical with imaginary parts of single-particle Green's
functions~(\ref{gf}).  When shifting the single-particle
energies by mean fields generally dependent on relative momenta
of~1 and~2, and~3 and~2, one precisely accounts for the change
in the density of relative states, in~feeding of the states.
Besides the
change of relative momenta due to mean field, that corresponds
to~$\Delta
\tau_1^{class}$, the~levels of~1 and~2 or~3 and~2 are pulled
apart
or pushed together when mean field is energy-dependent,
changing
the density, what corresponds to~$\Delta \tau_2^{class}$.
Then the rates calculated with single-particle energies in the
final density of states, would yield population of states
consistent with these first two times in~Eq.~(\ref{ergcl}),
if~the particles could leave these
states during the whole time such as obtained with
an~inclusion of the forward time delay or simply adoption of
mean field.

Ergodicity when many scatterings occur is
discussed in more detail in the~Appendix, together with the
single-particle spectral densities.  We~show there, that,
in~fact, population of different states could be made
consistent with an~ergodic
constraint involving all three delay times in~(\ref{ergcl}),
i.e.~also the~scattering delay, if~one
included in the Green's function~(\ref{gf}) the imaginary part
of the ${\cal T}$-matrix
and allowed for a~separate dependence of the matrix
on~energy and relative momentum.  Modification of the
scattering or transition rates in terms of such complete
spectral functions, nonetheless, might
not be practical except for fully equilibrium situations.
Also, it~needs to~be stated
that a~modification of the final-state density for
two particles such as~1 and~3 on account of scattering with
third particle, {\em independent} of a~modification of matrix
element for~1 and~3, might not be proper.
With regard to scattering, correlations could
be important, persisting throughout the interaction
process.

It will be seen in the examples of Secs.~\ref{Res}
and~\ref{NNScat} that the
contribution to the density of states from the mean field alone
can be very significant.
If~scattering is then done in the presence of strong mean
fields, it~is important that matrix elements get modified~too.
In~case of momentum-independent mean fields, this just amounts
to the calculation of scattering rates in terms of kinetic
particle energies only.  Situation can be more cumbersome in
case of momentum-dependent fields as one generally loses the
convenience of working in the rest frame of scattering
particles.  A~particularly strong
momentum-dependence of the mean field
results for pions in nuclear matter due
the~derivative coupling to
nucleons\cite{mig78,kor90}.

Rates and scattering prescriptions may be complicated when the
time delay is negative.  In~such a~case the prescription may
effectively exclude a~relative volume, e.g.~an~energy-dependent
hard core.  One~may not want to fill the excluded phase-space
volume in the scattering with a~third body and with this
populate relative
states following ergodic theorem including the scattering delay
time.  Scattering prescriptions are investigated in
Sec.~\ref{Simu}, including one with a~hard core prescription
and one where effective time delays are generated by
correlating the outgoing scattering angle with the impact
parameter.

\subsection{Manipulating Time Delays}
\label{manipulate}

If one solely aims at satisfying the ergodic constraint, and thus properly
reproducing thermodynamic properties of a~system, while ignoring the physical
differences behind the mean field or forward time delay and the delays for
scattered waves, then one can absorb all interaction effects, in~the lowest
nontrivial order in density, exclusively into the mean
field~or, alternatively,
into the delays for scattered particles.  We~shall illustrate our points by
considering the correction to the pressure in~(\ref{presde})
and~(\ref{deltap}).

Thus, from (\ref{delt}), (\ref{udef}) and~(\ref{clasum}),
it follows that the density of two-particle states in energy would
be properly reproduced when ignoring the delays for scattered
waves and, in~place of the mean field $u$ in the center
expression in~(\ref{udef}), using a~field $u'$ given
by~(\ref{udef}) with $\sin{2 \delta_{\ell}}$ replaced by~$2
(\delta_{\ell} (E) - \delta_{\ell}(0))$.  As~far as the
correction to the pressure is concerned, this corresponds
to rewriting the expression (\ref{deltap}) as
\begin{equation}
\Delta P  =  {1 \over 2}  \int d {\bf p}_1 \, d {\bf p}_2
\,  f (p_1) \,
f (p_2) \, \left\langle ( {\bf p}_1 - {\bf p}_2 )/
2  \right|  {\rm Re} \, {\cal T}'
 \left| ( {\bf p}_1 - {\bf p}_2 )/ 2
\right\rangle
,
\end{equation}
where
\begin{equation}
\langle p | {\rm Re} \, {\cal T}' | p \rangle = {2 \pi v
\over k^2}
\sum_{\ell} (2\ell+ 1)(\delta_{\ell} (0) - \delta_{\ell} (E)).
\end{equation}
The replacement of the sine by its argument in the field is
actually a~good approximation
when the phase shifts are low compared to $\pi /4$.
The general conclusion then is
that the delays for scattered waves are relatively
unimportant for the thermodynamic properties when the phase
shifts are low, no matter what are the values of these delays.

If the mean field were to be ignored, then, in~order to
properly reproduce the density of two-particle states,
from (\ref{delt}) and (\ref{tauc}), the mean delay time for
scattering should be taken equal to
\begin{equation}
\Delta \tau_{s}^{\prime} = {\sum_{\ell} (2 \ell + 1) \, {d
\delta_{\ell} \over dE} \over 2 \sum_{\ell} (2 \ell + 1)
\sin^2{\delta_{\ell}} } , \end{equation}
and for scattering in only one partial wave
\begin{equation}
\Delta \tau_{s}^{\prime} = \frac{1}{2\sin^2{\delta_{\ell}}}
\frac{d\delta_{\ell}}{dE} ,
\label{tauprime}
\end{equation}
in place of~(\ref{tauda}).
The correction to the pressure (\ref{deltap}) is then rewritten as
\begin{equation}
\Delta P  =   - T \, {1 \over 2} \int d {\bf p}_1 \, d {\bf p}_2
\,  f (p_1) \,
f (p_2) \,  \sigma  \, v \,
\Delta \tau_s' .
\end{equation}
Some advantage of an~approach with delays put into scattering is that particles
move with uniform velocities.  One~problem, though, is~that the
delay times~$\Delta \tau_s'$ diverge at thresholds, where phase
shifts go to zero linearly
in the momentum with the coefficient of proportionality being equal to the
negative of scattering length.  In~Sec.~\ref{Res} we shall
discuss
resonance scattering including, of interest for heavy-ion collisions,
pion-nucleon scattering, and in that context discuss more the different
possible prescriptions for time delays.  Now~we turn to
to the role of~statistics.

\section{Time Delays and Statistics}
\label{Stat}

For any finite energy the density of states in energy is increases by
symmetrization and decreased by antisymmetrization.  Corrections to
thermodynamic quantities, such as the free energy density or pressure, arise
within the second order in density even in the absence of interactions.
Symmetrization also affects scattering processes both in that
the outgoing
states may be Pauli-blocked or Bose-enhanced and in that the
scattering amplitude may
be internally modified\cite{bar92}.  In~the following,
we~consider pressure
in an~equilibrated many-body system in terms of the ${\cal
T}$-matrix.  In~a nonequilibrium system, for low scattering rates,
the~${\cal T}$-matrix approximation\cite{dan84}
in the single-particle equations of motion leads to the
Boltzmann equation with rates
corresponding to two-particle collisions, enhanced or reduced on account of
statistics of final states and with medium-modified amplitudes.
We~assess
implications for time delays and scattering processes
in~simulations, following from ergodicity.

The pressure in a~many-body system at a~given temperature $T$
and chemical potential~$\mu$ may be generally
represented~as\cite{fet71}
\begin{equation}
P(\mu, T)  =  P_0(\mu, T) + \Delta P(\mu, T) ,
\label{PDP}
\end{equation}
where
\begin{eqnarray}
\Delta P(\mu, T) & = &   {1 \over 2} \int
{d{\bf P}\over (2\pi)^3} \,
{d {\bf p} \over (2\pi)^3}
\,
d E_t
\, {1 \over {\rm e}^{(E_t - 2\mu)/T} - 1}
\,
\nonumber \\ & & \times
 \int^1_0 d\lambda \,
\left(-{1 \over \pi} \right){\rm Im} \, \langle
{\bf p} | {\cal V} \, G_{\lambda} ({\bf P}, E_t) \left( | {\bf
p} \rangle \pm |-{\bf p} \rangle \right) .
\label{DP}
\end{eqnarray}
Here ${\cal V}$ is the interaction, ${\bf p}$ and ${\bf P}$ are relative and
total momenta, respectively, $E_t$ is total energy, and $G_{\lambda}$ is the
two-particle Green's function within a~system with the interaction scaled down
by a~factor of $\lambda$.  The~negative of the imaginary part of this function,
divided by~$\pi$, plays a~role of the two-particle spectral function.  The
statistical factor in~(\ref{DP}) is bosonic, as is appropriate for a~state of
two particles with the same statistics.  The~apparent change in sign of the
correction to the pressure above in the classical limit ($\mu /T \rightarrow -
\infty$), compared to (\ref{presde}), is associated with the fact that there
the pressure is expressed as a~function of density while
in~(\ref{DP}) it is expressed as a~function of $\mu$.

  In the
${\cal T}$-matrix approximation\cite{dan84} the two-particle
Green's function satisfies
\begin{equation}
G = G_0 + G_0 \, {\cal V} \, G,
\label{GG0}
\end{equation}
and explicitly
\begin{eqnarray}
\langle{\bf k}|G ({\bf P},E_t)|{\bf k}_1\rangle & = &
\langle{\bf k}|G_0 ({\bf P},E_t) |{\bf k}_1\rangle + \int
{d {\bf k}_2 \over (2 \pi)^3} \,
{d {\bf k}_3 \over (2 \pi)^3} \,
\langle{\bf k}|G_0({\bf P},E_t)|{\bf k}_2\rangle \,
\nonumber \\ & & \times
\langle{\bf k}_2|{\cal V}|{\bf k}_3\rangle \,
\langle{\bf k}_3|G ({\bf P},E_t)|{\bf k}_1\rangle
\label{Green} .
\end{eqnarray}
The noninteracting Green's function in the above is equal to
\begin{equation}
\langle{\bf p}|G_0({\bf P},E_t)|{\bf p}'\rangle
= (2 \pi)^3 \delta({\bf p} - {\bf p}') \,
{N({\bf p}, {\bf P}) \over E_t - P^2/4m - p^2/m + i \epsilon} \, .
\label{G0}
\end{equation}
The factor of $N$ stems from an~equal-time commutator of the operators for two
particles,
\begin{eqnarray}
N({\bf k},{\bf P}) & = & (1 \pm f({\bf P}/2+ {\bf k})) (1
\pm f({\bf P}/2 - {\bf k})) - f({\bf P}/2+ {\bf k}) \, f({\bf
P}/2 - {\bf k}) \nonumber \\ & = & 1  \pm f({\bf P}/2+ {\bf
k}) \pm f({\bf P}/2- {\bf k}),
\label{NP}
\end{eqnarray}
and the upper signs refer to bosons and lower to fermions.
The ${\cal T}$-matrix and Green's function are related with
\begin{equation}
{\cal T} = {\cal V} + {\cal V} G {\cal V} ,
\end{equation}
and thus the ${\cal T}$-matrix satisfies
\begin{equation}
{\cal T} =
{\cal V} + {\cal V} G_0 {\cal T} .
\label{Gal}
\end{equation}
The factor of $N$ in $G_0$ in the above equation, on
considering the case of
a~resonance, can be related to the fact that, for a~bosonic
two-particle state, the~width is equal to the difference
between the decay and the formation rates\cite{dan84}.

On integrating in~(\ref{DP}), we obtain, from (\ref{GG0}),
\begin{eqnarray}
 \int^1_0 d\lambda
\,
{\rm Im}
\, \langle
{\bf p} | {\cal V} \, G_{\lambda}  \left( | {\bf
p} \rangle \pm |-{\bf p} \rangle \right)
& = & {\rm Im}
\, \langle
{\bf p} | \sum_{n=1}^{\infty} {1 \over n} \left( {\cal V} G_0
\right)^n
 \left( | {\bf p} \rangle \pm |-{\bf p} \rangle \right)
\nonumber \\
& = & - {\rm Im}
\, \langle
{\bf p} | \, \log{\left(1 - {\cal V} \,G_0 \right)}
 \left( | {\bf p} \rangle \pm |-{\bf p} \rangle \right)
\nonumber \\
& = &
 \langle
{\bf p} | \, {\rm atan} \left( {\cal V}  \,
(1 - {\cal V} \, {\rm Re} \, G_0 )^{-1} \, {\rm Im} \, G_0 \right)
 \left( | {\bf p} \rangle \pm |-{\bf p} \rangle \right)
\nonumber \\
& = &
 \langle
{\bf p} | \, {\rm atan} \left( {\cal R} \, {\rm Im} \, G_0
\right)
 \left( | {\bf p} \rangle \pm |-{\bf p} \rangle \right) ,
\label{DPr}
\end{eqnarray}
where ${\cal R}$ satisfies
\begin{equation}
{\cal R} = {\cal V} + {\cal V} \, {\rm Re} \, G_0 \, {\cal R}.
\end{equation}
On introducing then the matrix $\widetilde{\cal R}$ that is
hermitian in the spherical angle and may be diagonalized,
\begin{eqnarray}
\widetilde{\cal R}(p, \Omega, \Omega', {\bf P}) & = & {\rm sgn}
(E_t - 2\mu) \, \left| N(p,P,\Omega) \, N(p,P,\Omega')
\right|^{1/2} \,  \langle
p, \Omega | {\cal R} ({\bf P}, E_t)
 |  p, \Omega ' \rangle
 \nonumber \\
& = & \sum_n {\sl r}_n (p,P) \, {\cal Y}_n(p,P,\Omega) \, {\cal
Y}_n^* (p,P,\Omega'),
\end{eqnarray}
where $E_t = P^2/4m + p^2/m$, and
$\lbrace {\cal Y}_{n} \rbrace
$ form an~orthonormal set in spherical angle (with a~definite
symmetry under inversion), we can rewrite (\ref{DP}) as
\begin{eqnarray}
\Delta P(\mu, T) & = & {1 \over 2} \int
{d{\bf P}\over (2\pi)^3} \,
{d E \over \pi }
\, {1 \over {\rm e}^{(E + P^2/4m - 2\mu)/T} - 1} \, \sum_n
\delta_n (p, P) ,
\label{DPps}
\end{eqnarray}
where $\delta_n = {\rm atan} \left(-p^2 {\sl r}_n/8 \pi^2 v
\right)$.

In terms of the phase shifts $\delta_n$ introduced above, the on-shell
${\cal T}$-matrix can be expressed~as
\begin{eqnarray}
\langle p, \Omega | {\cal T} ({\bf P}, E_t) | p, \Omega'
\rangle & = & - {8 \pi^2 v \over p^2}
\, {\rm sgn}
(E_t - 2\mu) \, \left| N(p,P,\Omega) \, N(p,P,\Omega')
\right|^{-1/2}  \nonumber \\ & & \times \sum_n \sin{\delta_n}\,
{\rm e}^{i\delta_n} \, {\cal Y}_n(p,P,\Omega) \, {\cal
Y}_n^* (p,P,\Omega') .
\end{eqnarray}
On carrying partial integrations in~(\ref{DPps}),
the correction to the pressure may be decomposed into the
mean-field and scattering contributions that take the form, in
terms of the ${\cal T}$-matrix,
\begin{eqnarray}
\Delta P(\mu, T) & = & - {1 \over 2} \int {d
{\bf p}_1
\over (2\pi)^3} \int {d {\bf p}_2 \over (2\pi)^3} \,
f(p_1) \, f(p_2) \, {\rm Re} \, \langle {\bf p}
| {\cal T} ({\bf P}, e(p_1) +
e(p_2)) \left( |  {\bf p} \rangle  \pm | -  {\bf p} \rangle
\right)
\nonumber \\
 & & + {1 \over 2} \int {d {\bf P} \over (2 \pi)^3} \, {P^2 \over
6m} \int {d {\bf p} \over (2 \pi)^3 } \int d\Omega \Big( f(p_1)
\, f(p_2) \, {d \sigma \over d \Omega} \, v \Delta \tau_s \,
(1 \pm f(p_1')) (1 \pm f(p_2') ) \nonumber \\
& & -  f(p_1)
\, f(p_2) \, {d \sigma \over d \Omega} \, v \Delta \tau_s \,
f(p_1') \,f(p_2') \Big) ,
\label{presc}
\end{eqnarray}
where $f = (\rm{e}^{(e(p)-\mu)/T} \mp 1)^{-1}$, $d \sigma / d \Omega = (m/4\pi
)^2 | \langle {\bf p} |{\cal T} ({\bf P}, E_t) \left( | {\bf p}' \rangle \pm |
- {\bf p}' \rangle \right) |^2$, integration is over the spherical angle of
$2\pi$, and the time delay for scattering is
\begin{equation}
\Delta \tau_s =\Bigg({\partial \over \partial \left(p^2/m
\right)} - {\partial \over \partial \left( P^2/4m \right) } \Bigg) \,
{\rm Im} \left( \log{\langle {\bf p} | {\cal
T} ({\bf P}, E_t) \left( | {\bf p}' \rangle \pm | - {\bf p}'
\rangle \right)} \right)  .
\label{tauT}
\end{equation}

Following (\ref{PDP}) and (\ref{presc}), in~order to produce proper changes in
the pressure in simulations, associated with two-particle interactions, it~is
necessary to include the effects of mean field on particle motion, and to delay
the collision processes, according to the expressions in terms of ${\cal
T}$-matrix, with an~internal symmetrization and symmetrization with other
particles in the medium accounted for in the final and intermediate states
(Eqs.~(\ref{Gal}), (\ref{G0}), (\ref{presc}),
and~(\ref{tauT})).  Besides
delaying direct collisions, where two particles alter their momenta and spins,
by~$\Delta \tau_s$, it~is necessary to delay exchange collision processes,
where pairs of particles meet and interchange pairwise quantum numbers, by~$-
\Delta \tau_s$.  (The latter holds for particles of one statistics.
In the case of particles of opposite statistics, the exchange collisions should
be delayed by~$\Delta \tau_s$.)  The exchange collisions may be thought of as
processes\cite{fet71} where two particles collide and, while in an
intermediate 2p-2h state, encounter two more particles, ending up in the
interchange of the quantum numbers.  The~cross section,
from~(\ref{presc}), is~the same as for the direct collisions.
Notably, there is no room for the
exchange collisions in the Boltzmann equation that ignores the duration of
interactions, since these processes leave the occupations of single-particle
states unaltered.  These processes should, nonetheless, appear in a~possible
quantum Enskog equation, since their duration affects thermodynamic quantities.

\section{Resonance Scattering}
\label{Res}
\subsection{Sharp Breit-Wigner Resonance}
\label{Sharp}

An obvious example, that can serve to illustrate the time delays and different
prescriptions, is that involving a~sharp resonance.  If~the width of the
resonance $\Gamma$ is small compared to its energy $E_R$, then the phase shift
in the vicinity of $E_R$ is given by
\begin{equation}
\tan{\delta} = - \frac{\Gamma /2}{E-E_R} \, .
\label{tand}
\end{equation}
This yields the following time-delays for the scattered and forward-going
waves, respectively,
\begin{equation}
\Delta \tau_{s} = \frac{d\delta}{dE} =
\frac{\Gamma} {2[(E-E_R)^2 + \Gamma^2/4]} \, ,
\label{tausr}
\end{equation}
and
\begin{equation}
\Delta \tau_f \times {\pi B^2 \over \sigma}  = \Delta \tau_s' -
\Delta \tau_s = {(E - E_R)^2 - \Gamma^2/4
\over \Gamma [(E - E_R)^2 + \Gamma^2/4 ] } \, ,
\label{taufr}
\end{equation}
where $\Delta \tau_f$ is the time delay of the forward going wave packet with
total cross-sectional area~$\pi B^2 = ( \pi /k^2)\sum_{\ell}(2
\ell + 1)$.
Figure~\ref{restim} illustrates the different times.
It~is seen that the delay for
the forward-going wave turns negative in the vicinity of the
resonance.  For~light in a~dielectric medium\cite{jac75} this
corresponds to the increase in
group velocity for packets with resonant frequencies.

The energy-averaged delay for the forward-going wave~$\Delta
\tau_f$ is zero.
(This may be expected whenever phase-shift variation is limited to a~narrow
range in energy.)  Weighted with the cross section, the~average
delay time for
scattered waves over energy is equal to the inverse width
coinciding with naive expectations.  The~actual delay time
for scattered waves~(\ref{tausr}) is twice as high at the
resonance, see~also\cite{mes66}, and~it drops rapidly with
energy when going away from the vicinity of that resonance.

Time delay equal to the inverse width is further obtained when
putting all delays into scattering, from~(\ref{tausr})
and~(\ref{taufr}),
\begin{equation}
\Delta\tau_s'=\frac{1}{\Gamma} \, .
\end{equation}
It is now a~common prescription in relativistic simulations to
use an~energy-independent delay given by the inverse width.
Provided that particles are not also propagated through a~mean
field and resonances are indeed of a~Breit-Wigner form, such
a~prescription would yield correct thermodynamic functions up
to the second order in density.

\subsection{$\pi N \Delta$ System}

Different resonances in a~$\pi N$ system may be identified as
independent
particles.  For~the lowest~$P_{33}$ $\Delta$-resonance,
the~width is comparable
to the energy of the resonance above the $\pi N$ threshold and, as such, this
width exhibits a~significant energy variation.  Interacting system of pions,
nucleons, and deltas, is of considerable interest for heavy-ion collisions in
the beam-energy range from few hundred~MeV/nucleon to
few~GeV/nucleon.

In this section we address three issues.  First, by~studying
$\pi N$ phase shifts
we~calculate time delays and indicate differences with the
Breit-Wigner case
above.  Second, we~point out that a~quantum decomposition of
the effective
increase of the~density of states represented by the time delay
would include
contributions from {\em both} the~$\Delta$ and~$\pi N$
components.  Finally, we~point out a~practical difficulty in
separating the~$\Delta$ and~$\pi N$ components.

The time delays can be calculated given the experimentally
determined $\pi N$~phase shifts\cite{rowe78,hey71}.  The~delay
of the forward wave (divided
by the fraction of particles that scatter as in~(\ref{taufr}))
and the delay of the scattered
wave are shown for a~$\pi^+ p$ system in~Fig.~\ref{deltatimes}.
In~many prescriptions, time delays are
only incorporated into scattering events, which would mean that
the combination of
the two contributions~$\Delta \tau_s'$ is~appropriate.
One~should note that this combination differs
significantly from the Breit-Wigner result and is
extremely divergent at threshold due to a~rapidly declining
cross section, cf.~Sec.~\ref{manipulate}.  If~the mean field is
consistently
incorporated into a~simulation, the~appropriate delay
is~$\Delta \tau_s$ which behaves as
the derivative of the phase shift with respect to energy which
is approximately
the scattering length divided by the velocity in the vicinity
of the threshold.  This threshold divergence is much weaker
than that of~$\Delta \tau_s'$.

The imaginary part of a~$\Delta$ Green's function in a~thermal
system gives the number of states for the resonance per unit
energy and volume, $\Delta \rho '' = - {\rm Im} \,
g_{\Delta}/\pi$\cite{kad62,dan84}.  The~requirement then of
a~consistency with the~number of colliding~$\pi N$ pairs gives
the time~$\Delta \tau_s ''$ during which the~$\pi N$ pair
should turn into the resonant particle,
\begin{equation}
{ \Delta   \rho '' (E) \over \rho_0
(E)} = {\sigma (E) \, v \, \Delta \tau_{s } '' \over V } \, ,
\end{equation}
with a result that is the inverse of resonance
width\cite{cug82,dan91},
\begin{equation}
\Delta \tau_{s } '' = {1 \over \Gamma (E) } \, .
\label{1G}
\end{equation}
This time differs from~$\Delta \tau_s'$, such as
in~Eq.~(\ref{tauprime}), that would, in~particular, involve the
energy derivative of~$\Gamma$; likewise~$\Delta \rho '' $
differs from~$\Delta \rho$.  Interestingly, within any single
spin-isospin channel, either correction to
the density of states, $\Delta \rho$ or $\Delta \rho ''$,
integrates over energy to unity, i.e.~one net state is gained.
For sharp resonances, with couplings and widths independent of
energy, there is no difference between~$\Delta \tau_s'$
and~$\Delta \tau_s ''$, as is apparent from Sec.~\ref{Sharp},
and there is no difference between~$\Delta \rho$ and~$\Delta
\rho ''$.

On studying the density of states for pions or nucleons as in
the Appendix, one finds a~change, per~$\pi N$ pair, that is
equal to~$\Delta \rho (E) - \Delta \rho '' (E)$,
where $\Delta \rho = (1/\pi) (d \delta / dE)$ in any single
channel, i.e.~one that precisely compensates the discrepancy above.
Same type of discrepancy and compensation is found when
applying the considerations to a~system of pions and rho
mesons, where $\pi + \pi \leftrightarrow \rho$ reactions take
place.  If~one were to ask about, in the last system, how
many $\rho$~mesons decay
into dilepton pairs, the~answer would involve the density of
rho states within the system, rather than the overall change
in the density of states associated with the resonance
formation,
important for thermodynamic considerations.  For~certain
questions one~has to keep in mind that the time delays derived
before correspond to the change in the overall density of
states and not necessarily to the existence of the resonant
particles.

The time for the conversion into a~resonant particle in
scattering~(\ref{1G}) may diverge strongly when
threshold is approached, which parallels the situation when the overall
delay time associated with the~interaction is forced
onto the scattering. (The~conversion time for
a~spherical wave, that should be identified as $\Delta \tau_{sph} =
2 \pi \Delta \rho''
\propto \Gamma (E)$, tends to zero at the threshold.)  In~practice,
manipulations of the conversion time, dividing this time
between the scattered and forward waves, may
pose more difficulty than the manipulations of the overall
delay time, as~negative conversion times
cannot be simulated.

\subsection{$s$-Channel Scattering of Massless Partons}

An example, where time delays associated with interaction are
relevant, is~the
collision of partons in ultrarelativistic nuclear reactions.  Partons are
copiously produced early on in reactions and the goal of simulating partonic
cascades\cite{gei92,wan91,esk93} is to determine the equilibration time scale
and initial equilibrated energy density.  Most partons at midrapidity are
produced far off shell and decay via bremsstrahlung.  Thus, quantum
considerations are necessary to establish the duration of processes.

While we shall not tackle the general problem here, we shall
try to gain
insight by considering the simplified example of elastic
scattering of two
massless partons in the $s$-channel, assuming that the
intermediate particle is
massless.  This turns out to be quite similar to the case
of~$\pi N$ scattering
close to the threshold.  Since there are no energy scales, to
lowest order in
perturbation theory the phase shift must depend only on the
coupling constant~$\alpha$,
\begin{equation}
\tan{\delta} = - c \alpha .
\label{tanp}
\end{equation}
Then, the overall time delay for a~spherical wave from
(\ref{tanp}) is zero.  This
peculiar result will arise from any theory with no energy scale
since phase shifts are dimensionless.

However, QCD acquires a~scale~$\Lambda$ through
renormalization that gives the coupling constant an~energy
dependence\cite{fie89}:
\begin{eqnarray}
\tan{\delta} = c\alpha(E)
=  c \, {\frac {4\pi} {\beta_0 \,\log{(E^2 / \Lambda^2)}}} \, .
\end{eqnarray}
With this, the time delay for
spherical waves becomes
\begin{eqnarray}
\Delta \tau_{sph} = 2 \, \frac{d\delta}{dE}
= \frac{\beta_0}{c\pi E} \, \sin^2{\delta}
= \frac{\beta_0}{c\pi E}
\, \frac{c^2\alpha(E)^2}{1+c^2\alpha(E)^2} \, .
\label{deltp}
\end{eqnarray}
If all the time delay is put into the scattered wave, as in
Sec.~\ref{manipulate}, then the correct time delay for the
scattered wave acquires a~particularly simple form
\begin{equation}
\Delta \tau_{s}' = \frac{\beta_0}{4 \pi c E} \, .
\end{equation}
Note that the time delay does not involve the coupling constant
and behaves
as inverse energy.  If~one chose the time delay equal to the inverse width
for the intermediate state, then would certainly have obtained
a~time proportional to~$1/(\alpha E)$, as~the width would be
proportional~to~$\alpha$.

Since an~$s$-channel scattering involves an~intermediate state
very far from
being on-shell, the questions involving the time delay for such
a~process may
not be so crucial since these processes are rather rapid.
Of~a~greater
concern is the formation of partons through bremsstrahlung
involving intermediate states which are nearly on-shell.  Since
such processes create the
majority of soft particles in an~ultrarelativistic $pp$
collision, the~issue of
when and where such particles appear can greatly affect
estimates of the
initial thermalized energy density.  Unfortunately, such two to
three or more particle
processes are outside the scope of this analysis, but similar
problems have
been addressed in the context of decaying hadronic
resonances\cite{ber94}.

\section{Nucleon-Nucleon Interaction}
\label{NNScat}
The nucleon-nucleon system is one for which the quantum-mechanical
scattering-amplitudes have been most carefully measured in
physics.  It~represents the most relevant case of scattering
for heavy-ion physics, where
semiclassical simulations utilizing single-nucleon degrees of freedom are
commonly used to model heavy-ion reactions.  At moderate densities and high
temperatures, when cluster formation is unlikely, it~may be reasonable to
assume that nucleons interact two at a~time.  Using phase shifts
inferred from scattering data, one can calculate quantum mechanical scattering
amplitudes and determine the appropriate time delays following the
prescriptions outlined in the previous sections.

First we present the calculation of the time delay time for the scattered wave
at a~given angle, $\Delta \tau_s (\theta)$, when averaging over the spin and
isospin directions.  Due to antisymmetrization and the conservation of angular
momentum, isospin and parity, the magnitude of the net nucleon spin is
conserved in $NN$ interactions.  The~$S=0$ amplitude for a~given isospin is
given by
\begin{equation}
f^{S=0} (\theta) = - {1 \over k} \sum_{\ell ,j} (2 \ell + 1)
P_{\ell}( \cos{\theta}) \, {\cal T}_{\ell} ,
\label{fS0}
\end{equation}
where only even or odd values of $\ell$ are included.  That is compensated by
the factor before the sum in~(\ref{fS0}) being twice as large as in
(\ref{fteta}).  Given the conservation laws, the $S=1$ amplitude is of the
general form,
\begin{equation}
f^{S=1}_{\mu \mu'} (\theta , \phi) = - {1 \over k} \sum_{\ell}
\sqrt{4 \pi (2 \ell + 1 )} \, \langle \ell \, 0 \, 1 \, \mu' |
j \mu \rangle  \,
\langle \ell' \, (\mu - \mu') \, 1 \, \mu' | j \mu \rangle \,
Y_{\ell'
\, \mu - \mu'} (\theta, \phi) \, {\cal T}_{\ell \ell' j} \, ,
\end{equation}
where ${\cal T}_{\ell \ell' j} = i (S_{\ell \ell' j} -
\delta_{\ell
\ell'})$, and for the coupled waves diagonal matrix elements
are $S_{\ell \ell} = \cos{2 \overline{\epsilon}_J} \exp{2 i
\overline{\delta}_{j \ell}}$, and off-diagonal elements $S_{\ell \ell' j} = i
\sin{ 2 \overline{\epsilon}_J}
 \exp{ i( \overline{\delta}_{j \ell}+
\overline{\delta}_{j
\ell'})} $, where $\overline{\epsilon}$ is mixing parameter.
In carrying calculations up to a~laboratory energy of $E_{lab}
= 400$ MeV, we use all partial waves with both $\ell$ and $j$
less than 5.
The mixing of waves is not very strong in this region.
The phase shifts are generated using a~potential model that had
been carefully fitted to describe the $NN$ data\cite{sto84}.

Using Eq.~(\ref{tausa}), one obtains a~delay time $\Delta
\tau_s$ that depends on the scattering angle.  This time is
shown by different lines in Fig.~\ref{tausNN} for several values of $E_{lab}$.
Due to the averaging over initial spin and isospin and the amplitude
antisymmetrization, the time is symmetric with respect
to~90$^{\circ}$.  The~time, averaged over angles, is
additionally shown as a~function of the
laboratory energy by a~solid line in Fig.~\ref{tauNN}, together with the
forward delay time from Eq.~(\ref{taufa}).  The~observed negative delay times
for scattered waves reflect the negative derivatives of $S$-wave and other
phase-shifts with respect to energy, weighted in the average time with
contributions to the cross section.  The~falling phase-shifts make the
interaction, with regard to scattering, effectively repulsive above $E_{lab}
\sim 2.5$ MeV.  Notably, the $NN$ interaction is often {\em only} considered
repulsive when phase shifts are predominantly negative, although the energy
derivatives of the phase shifts also necessitate a~consideration. Both the $T =
0$ and $T=1$ phase shifts fall with energy, as the Levinson's theorem requires
that the deuteron formation in $T=0$ channel and enhancement of the density of
states in the $d^*$ region in $T=1$ channel be compensated by a~depletion in
states at higher energies.
In a~semiclassical consideration, the delay time for scattering
should be limited from below by~$-2 d /v$, where $ d $ is
interaction range and $v$ is relative velocity.  While the
negative delay times in Fig.~\ref{tauNN} decrease in magnitude
with an increasing
laboratory energy, their decrease is faster than implied by the
above limit.  At very low energies, the times for the scattered
waves become positive.  At $E_{lab} {\buildrel < \over \sim}
100$~keV they
begin to be governed by the singlet scattering length, $\Delta
\tau_s \simeq - a_s /v$.

For~comparison of angular dependence,
the delay time for a~hard-sphere repulsive-scattering
(discussed more in the next section) as a~function of the
scattering angle is
\begin{equation}
\Delta \tau_s (\theta) = - {2  d  \over v} \sin{\theta \over 2}
,
\label{tausth}
\end{equation}
and averaged over forward and backward directions, given
the constant scattering cross section,
\begin{equation}
\overline{ \Delta \tau_s } (\theta) =
{1 \over 2} \big( \Delta { \tau}_s (\theta) + \Delta { \tau}_s
(\pi - \theta) \big) =
- {\sqrt{2}  d  \over v}
\cos{\left({\pi \over 4} - {\theta \over 2} \right)}.
\end{equation}
Correspondingly, for repulsive scattering, more negative delay times might be
expected at $\theta = 90^{\circ}$ than at $\theta = 0$.  Indeed, that is
observed in $NN$ scattering at the higher laboratory energies.  In~fact, in~the
periphery the interaction might be expected attractive and the delay times
might be expected to turn to zero or even positive in the forward directions,
and not just decline in magnitude as for hard-sphere scattering, and this can
be seen for the 360 MeV scattering in Fig.~\ref{tausNN}.
In~passing, let us note that to
properly isolate forward and backward directions, constructing an~amplitude
prior to the antisymmetrization, one should continue the phase shifts over the
missing partial waves.  If~one were to identify a~hard-sphere radius $ d $ for
simulations from the delay at $90^{\circ}$, quite low values would have been
obtained compared to what was used in simulations\cite{sch91}, declining from
$ d = 0.60 - 0.80$ fm for $E_{lab}$ within the range $10-30$ MeV to $ d = 0.15
- 0.30$ fm for $E_{lab}$ within the range $100 - 400$~MeV.
At~low energies,
the times become more negative at $0^{\circ}$ than
at~$90^{\circ}$, and this is
due to $^3S_1$-$^3D_1$ interference.

The forward delay times are positive in a~wide energy range $3
{\buildrel < \over \sim} E_{lab}
{\buildrel < \over \sim} 150$~MeV, see~Fig.~\ref{tauNN}.
In~the lower portion of the range, this is due
to the fact that, for large $S$-wave phase-shifts, the real forward amplitude,
multiplied by a~relative momentum, increases, although the $S$-wave
phase-shifts decrease.  In~the higher portion of the above energy range, when
$S$-wave contributions are low, the forward delay times are positive due to the
positive energy-derivatives of phase shifts for some high partial waves which,
unlike in the delays for scattered waves, are not weighted by partial cross
sections in the forward direction.

Of some interest is the issue of elastic $NN$ interactions at very high
energies when amplitudes are primarily diffractive\cite{kah94}.
Schematically, a~purely diffractive amplitude may be represented, given
$S_{\ell} = 0$ for $\ell
\le \ell_c$ and $S_{\ell} = 1$ for $\ell > \ell_c$, as
\begin{equation}
f(\theta) = {i \over 2k} \sum_{\ell = 0}^{\ell_c} \, (2\ell + 1)
\, P_{\ell}(\cos{\theta}) .
\end{equation}
As a~diffractive amplitude is purely imaginary, its phase does not depend on
energy and the time delays for the elastically scattered waves identically
vanish, cf.~(\ref{tausa}).  Likewise, the~times for the forward
waves
(\ref{taufa}) vanish.  This is consistent with the concept of
particles moving freely around the interaction region.

We conclude this section with a~presentation of the pressure in a~low density
nuclear matter at moderate temperatures, as a~function of the density and
temperature, such as should be, generally, reproduced in simulations.  Within
the second order in density the contributions to pressure (beyond the free-gas
term $P_0 = n_B\,T$) come from the nucleon antisymmetrization, formation of
deuterons, and nucleon-nucleon scattering,
\begin{equation}
P = n_B \, T \left( 1 + a_2(T) \, n_B  \left( {2 \pi
\over m T } \right)^{3/2} \right),
\end{equation}
where the three respective contributions to the virial coefficient $a_2(T)$ are
given by
\begin{equation}
2^{5/2} \, a_2 (T) = {1 \over 4} - 3\,{\rm e}^{B_d/T} - \sum_{T j
\ell S} (2T + 1) (2j + 1) \int {dE \over 2 \pi} \, 2 \, {d
\delta_{j \ell}^{T S} \over dE} \, {\rm e}^{-E/T} ,
\label{a2}
\end{equation}
and $B_d$ is the deuteron binding energy.  The~$NN$ antisymmetrization
increases the pressure.  Deuteron formation lowers the pressure, while
scattering, at most temperatures, increases the pressure.  The~overall effect
of interactions, following the Levinson's theorem, declines with the increase
of temperature.  The~second virial coefficient (\ref{a2}) is shown as a
function of temperature in~Fig.~\ref{virial}.

\section{Delays in Simulations}
\label{Simu}
\subsection{Effect of Scattering Prescriptions}

Without explicitly delaying or advancing particles as they pass near each other
within a~simulation, the time spent by particles in the vicinity of one another
can be affected by the scattering prescription.  The~prescription can make
$\Delta \tau_s$ positive or negative.  Effects of scattering prescriptions on
macroscopic features of reaction dynamics at Bevalac energies have been
investigated in Refs.\cite{hal81,gyu82,mal84,sch91}, see also\cite{her95}.
We analyze two examples, hard sphere scattering and scattering at the point of
closest approach.

Our first example is that of hard-sphere two-particle scattering.  Given a
volume in relative coordinates of radius $R$, and a~hard-sphere potential that
rises as particles are distance $d$ away, the expected reduction in the density
of two-particle states within the relative volume, due to scattering, becomes
$\Delta \rho / \rho_0 = \Delta \tau^{coll} /
\tau_0 = - d^3/R^3$. Particles that make contact along a~line at an~angle
$\alpha$ to the direction of original relative motion, get deflected by an
angle $\theta = \pi - 2 \alpha$.  When reaching the boundary of the relative
volume defined with the radius $R$, the particles traverse a~relative distance
that is altered by the scattering.  The~alteration divided by the relative
velocity, gives the change in the time spent in the vicinity of the other
particle, $\Delta \tau_s (b)$, that depends on impact parameter and can be
worked out from geometry,
\begin{equation}
\Delta \tau_s (b) = - {2 \sqrt{d^2 - b^2} \over v} = - {2 d
\cos{\alpha} \over v} = - {2 d \over v} \sin{ \theta \over 2} .
\label{tausb}
\end{equation}
The average change in time due to scattering is then given an~integral over the
product of the probability density that a~specific scattering occurred,
 times the
change in time in that scattering,
\begin{equation}
\Delta \tau^{coll} = {\int db \, b \, \Delta \tau_s (b) \over
\int db \, b } = - {d^2 \over R^2} \, {4 \over 3} {d \over v} .
\label{tausc}
\end{equation}
In the last expression, the factors $d^2/R^2$ and $(4/3) \, d/v$ represent,
respectively, the probability that a~collision occurs and the average time lost
then in a~collision.  The~average time spent within the volume of radius $R$
follows from dividing the volume by cross-sectional area and velocity $\tau_0 =
(4 \pi/3) R^3 / \pi R^2 v = (4/3) R/v$, and we find that $\Delta \tau^{coll} /
\tau_0 = - d^3 /R^3$, as we expected. The pressure corresponding to
(\ref{tausth}), (\ref{tausb}) and (\ref{tausc}), in~absence of statistical
effects, from (\ref{presde}) and (\ref{deltap}), is
\begin{equation}
P = nT + n^2 \, T \, {2 \pi d^3 \over 3} \, .
\end{equation}
Positive delay times are obtained when scattering the
particles as if off a~thin
spherical shell of size $d$ open in the direction of motion (case of a~concave
rather than convex mirror).  Then expressions (\ref{tausb}) and (\ref{tausc})
remain valid but with changed signs.

Within most common prescriptions for scattering in simulations, it~is assumed
that particles come abreast of each other while at a~distance $b < d =
\sqrt{\sigma/\pi}$.  When deflected in a~direction making an~angle $\theta$
relative to the original direction, within a~plane at an~azimuthal angle $\phi$
with respect to the original plane containing particle trajectories, the
particles reach a~distance $R \gg d$ after a~time longer by
\begin{equation}
\Delta \tau_s (b,\Omega) = - {b \over v} \, \sin{\theta}
\cos{\phi} \, ,
\end{equation}
than in the absence of scattering.  If~the scattering is
repulsive ($\phi = 0$) and isotropic,
i.e.~$\cos{\theta}$-distribution is flat, then the average
delay time for a~given impact parameter becomes equal~to
\begin{equation}
\Delta \tau_s (b) = - {\pi \over 2} \, {b \over v} \, .
\end{equation}
By averaging over all impact parameters one further gets
\begin{equation}
\Delta \tau^{coll} = - {d^2 \over R^2} \, {\pi \over 3} \, {d
\over v} \, .
\end{equation}
Notably, the time lost in any one collision is reduced here by
only a~factor
$\pi/4$ compared to hard-sphere scattering.  By choosing $\phi = \pi$ in
collisions, one can produce the positive delay times.

Generally, given required delay times such that $|\Delta
\tau_s| < (2/3)
\, d \, \langle \sin{\theta} \rangle = (\Delta \tau_s)_{max}$,
these times may be generated making a~fraction $\nu$ of all
scatterings repulsive and a~fraction $1 - \nu$ attractive.
This fraction is given by
\begin{equation}
\nu = {1 \over 2} \left( 1 - {\Delta \tau_s \over (\Delta
\tau_s )_{max} } \right) .
\end{equation}

\subsection{Numerical Determination of Equations of State and
Transport Coefficients}

While we have limited ourselves to the discussion of effects of two-particle
interactions on thermodynamic properties, many-body calculations can provide
information accounting for interactions of a~few particles at a~time
\cite{ter87,fri81}.  The~prescriptions for interactions in simulations may
affect microscopic thermodynamic quantities within a~higher order than the
second in density and they can also affect transport coefficients.  Generally,
given the prescriptions, thermodynamic quantities within simulations and
coefficients can be determined numerically and confronted with those from
fundamental calculations.

All thermodynamic quantities can be derived once one knows the
pressure as a~function of the chemical potential and
temperature, $P(\mu ,T)$.  To determine
the pressure, the system may be enclosed in a~box of
macroscopic volume~$V$, in~contact with a~free non-interacting
gas with of chemical potential~$\mu$ and
temperature~$T$, possibly only within some external potential
lower than~$\mu$.
The contact with the gas can be made through the walls in one direction, and in
two remaining directions periodic conditions may be used.  If~clusters are
produced within the simulation, then the interfaces to the free gas can be made
impermeable to those.  Nucleons not in a~cluster, on the other hand, getting
into the free zone, would never return.  At the same time, nucleons from the
free zone with equilibrium phase-space distribution for
given~$\mu$ and~$T$
with the inclusion of statistics, would pour in into the box.  The~pressure
within the box could be then computed using the following virial-type
expression with terms for different possible ways of accounting for
interactions in a~simulation
\begin{eqnarray}
\nonumber
P & = & {1 \over \tau V} \Bigg\lbrace {1 \over 3} \Bigg(  \int dt
 \sum_j {\bf p}_j \cdot {\bf v}_j - \sum_{i < j}
 {\bf p}_{ij} \cdot {\bf v}_{ij} \, \Delta \tau_f^{ij}
+ \int dt  \sum_{i < j} {\bf F}_{ij} \cdot {\bf r}_{ij}
\\
& &
\hspace{4em}
+ \sum_{i < j} \Delta {\bf p}_{ij} \cdot {\bf r}_{ij}
\Bigg) + \int dt \, \int dV \, \left( \rho \, U -
\int_0^{\rho(t)} d\rho' \, U \right) \Bigg\rbrace \, .
\label{vir}
\end{eqnarray}
The above equation is limited to the mean field being momentum independent.

With regard to (\ref{vir}), pressure is, generally, recognized as the density
of momentum flux in equilibrium, in~any one direction.  In~(\ref{vir}) the
pressure is evaluated by taking a~trace of the momentum flux tensor and
dividing it by 3 for the three directions; $\tau$~is the time
over which the system is investigated.  The~first term on the
r.h.s.~of~(\ref{vir}), with a~sum over particles in a~box,
accounts for the transport of momentum when
particles move.  The~second term, with a~sum over particle encounters, accounts
for the situations when particles pass in the vicinity of one another and their
relative motion is delayed in a~simulation by~$\Delta
\tau_f$.  The~vector ${\bf p}_{ij} = ({\bf p}_i -
{\bf p}_j)/2$ is relative momentum and ${\bf v}_{ij} = {\bf v}_i - {\bf v}_j$
is relative velocity.  The~third term on the r.h.s.~of~(\ref{vir}) accounts for
transport of momentum in two-particle interactions treated
explicitly.  The~force ${\bf F}_{ij} = \dot{\bf p}_{ij}$ is
that due to particle~$j$ on particle~$i$.  Due to the
interaction, the relative momentum changes over a~distance
${\bf r}_{ij}$.  The~fourth term accounts for instantaneous changes of relative
momentum in collisions, and the final term accounts for the effects of
interactions treated in the mean-field approximation.  It~is
apparent in~(\ref{vir}) that positive forward delay times, for
a~given particle number,
reduce the pressure, and negative enhance.  Further, the attractive scattering
style, $\Delta {\bf p}_{ij} \cdot {\bf r}_{ij} < 0$, reduces the pressure, and
repulsive, $\Delta {\bf p}_{ij} \cdot {\bf r}_{ij} > 0$, enhances.

Transport coefficients such as shear viscosity or heat conductivity may be
determined within a~simulation by imposing different conditions within the free
gas beyond the two walls of the box in contact with that gas.  Provided that
the walls are perpendicular to the $x$-axis and at $x = \pm
\Delta x$, the~viscosity coefficient may be e.g.~determined by
giving to the free gas the
velocities in the $y$-direction equal to $\pm \Delta v$, respectively, in~the
two separated regions.  The~coefficient then follows from an~off-diagonal term
of the momentum flux tensor in the box
\begin{eqnarray}
\nonumber
\eta & = & - { 1 \over \tau \, V \, \Delta v / \Delta x }
\Bigg(  \int dt
 \sum_j { p}_j^y \, { v}_j^x - \sum_{i < j}
 { p}_{ij}^y \, { v}_{ij}^x \, \Delta \tau_f^{ij}
\\ & &
\hspace{7em}
+
\int dt  \sum_{i < j} F_{ij}^y \, x_{ij}
+ \sum_{ i < j} \Delta
p_{ij}^y \, x_{ij} \Bigg) \, .
\label{eta}
\end{eqnarray}
To gain an~insight into~(\ref{eta}), one may consider a~simple
assessment of the viscosity in a~medium when ignoring the
effects of the finite range and duration of interactions,
i.e.~investigating, in~particular, only the effects associated
with
the first term in~(\ref{eta}).  Provided that particles
propagate
freely between collisions for an~average time $\tau_F \approx
1/n \sigma \langle v \rangle$, a~particle at a~position $x_i$
of velocity $v_i^x$ would
have, on the average, a~momentum in the $y$-direction such as
characteristic for a~position this particle had
a time $\tau_F$ earlier.  With this, Eq.~(\ref{eta}) gives
\begin{eqnarray}
\nonumber
\eta & \approx & - { 1 \over V \, \Delta v / \Delta x }
\sum_i m \, \big\langle v^y \big\rangle (x_i - v_i^x \, \tau_F ) \,
v_i^x \\
\nonumber
& = &  - { 1 \over V \, \Delta v / \Delta x }
\sum_i m \, (x_i - v_i^x \, \tau_F ) \, (\Delta v/ \Delta x) \,
v_i^x    \\
& \approx & m \, n \, \langle (v^x)^2 \rangle \, \tau_F = {m \,
n
\, \langle v^2 \rangle \, \tau_F \over 3} \approx {m \, \langle
v \rangle \over 3 \sigma} \, .
\end{eqnarray}
The hard-sphere scattering is known to enhance viscosity to within the lowest
order in density, but the enhancement factor for viscosity is smaller than
that for pressure.  For~a~given shear then the medium with
hard-sphere
scattering, to a~lowest order, behaves as less viscous compared to one where
effects of interaction range and duration may be ignored.  For the numerical
determination of a~heat conduction coefficient, given
assumptions of a~simulation, the temperature should be set
different in the two regions with
free gas adjacent to the box where interactions take place.

\section{Conclusions}
\label{Conclu}

The principal goal of simulating heavy-ion collisions is to
infer the equation
of state of nuclear matter.  The~effective equation of state
for a~simulation
depends on several aspects: treatment of the mean field in
simulation, time delays for interactions and scattering
prescriptions, inclusion of various resonances.
In~this paper we have carried out a~detailed investigation of
the time
delays and of scattering prescriptions, and have shown that
they should be incorporated in a~coordination with the mean
field.  If~that is not followed, the~effective equation of
state may be inconsistent
with two-body
scattering which constrains thermodynamic quantities to within
the second order in the virial expansion.

In Sec.~\ref{Time} expressions were derived for the average
delay of an
outgoing scattered wave as a~function of the scattering angle.
In~Sec.~\ref{Ergo}, such delays in a~dynamics were shown
to be consistent with a~two-body density of states if,
besides, a~forward delay or mean field were included and
calculated in terms of the forward scattering
amplitude.  Other ergodically consistent prescriptions were
presented where all the effective
time delays were incorporated either entirely into the
scattering or into the mean field.
Alterations to these considerations for a~Fermi-degenerate
system were shown to be nontrivial in~Sec.~\ref{Stat}.
Sections~\ref{Res} and~\ref{NNScat} illustrated the
time-delay considerations with the examples of resonance and
$NN$ scattering.  In~Sec.~\ref{Simu} it was shown that
repulsive and attractive scattering schemes can be
interpreted in terms of time delays and equations of state
associated with these schemes may be understood
quantitatively at a~two-body level.  The implications of
scattering prescriptions for transport coefficients were also
discussed and a~practical method of determining the equation of
state and coefficients for a~simulation was presented.

We conclude by giving some perspective to the considerations
discussed here.
Most of the flow in high-energy ${\buildrel > \over \sim}
1$~GeV/nucleon heavy-ion collisions stems from
the one-body free gas pressure.  The~interactions of particles
affect the pressure and can increase the observed flow of
a~collision by~$\approx 50\%$ compared to a~free gas.  As~experiments are now
able to measure sidewards flow and squeeze-out flow differences
to better than~20\%, a~detailed
and careful understanding of the simulations becomes crucial.  Demonstrating
that a~simulation reproduces experimental results to within
(10--20)\% has little
meaning unless thermodynamic properties of the simulation are understood to
within (10--20)\% as well.  Simulations are becoming more
sophisticated.  Important
aspects, such as e.g.~the change of the dispersion relation
for pions, are being
incorporated through energy-dependent mean fields.
At very relativistic
energy, simulations must incorporate a~large number of
resonances which can be
very broad compared to the temperature. It~is hoped that the
prescriptions and
constraints presented here will contribute to~both the
development of improved
codes and to a~better understanding of existing approaches.

\acknowledgements
The authors are indebted to Harry Lee who generously generated the $NN$ phase
shifts used in the analysis.  They further thank M.~Herrmann,
G.~F.~Bertsch, E.~Shuryak, and S.~Ayik for discussions that
contributed to this paper.  P.D.~thanks for the hospitality
extended to him at the Institute for Nuclear Theory
in Seattle, where part of this work was carried.  This work was partially
supported the National Science Foundation under Grant Nos.~PHY-9403666 and
PHY-\ldots and by the Department of Energy under Grant No.~FG06-90ER40561.

\appendix
\section*{}
In this Appendix we investigate changes
in two-particle scattering-rate induced by
the~presence of other particles,
within the lowest order in density,
in~the context of ergodic
theorem.  To~reach directly essential results we~adopt
simplifying assumptions.
First, we~concentrate on a~subsystem of
one of the particles engaged in scattering, denoted as~1, and
the spectator particle denoted as~2.  We~assume that these
two particles interact only perturbatively with all other
particles
within the system, but~not necessarily with one another.
If~the system has many particles confined to a~finite
volume and sampling is done over a~long time~$\tau$, then we
may write
an~ergodic theorem for the subsystem of particles~1 and~2
alone:
\begin{equation}
{ 1 \over \tau} \,  {d\tau \over d{\bf p} \,
dE }
=
\left(
\int dE' \,
{d{\sl n} \over   dE' }
\, {\rm e}^{- E '/T} \right)^{-1} \,
{d{\sl n} \over  d{\bf p} \, dE }
\, {\rm e}^{- E/T}
\, .
\label{dtaun}
\end{equation}
In the above ${d\tau / d{\bf p} \,
dE }$ denotes time spent in a~given set of internal
states, per unit volume of energy-momentum space, ${d{\sl
n} / d{\bf p} \,
dE }$ denotes the number of states, and~$
{\rm e}^{- E/T}$ is relative probability that any single
internal state is occupied at any instant.
An~equivalent formulation of the ergodic theorem~(\ref{dtaun})
is that for any two states~$({\bf p},E)$ and~$({\bf p}',E')$:
\begin{equation}
\left( {d\tau \over d{\bf p} \,
dE } \right)^{-1} \,
{d{\sl n} \over  d{\bf p} \, dE }
\, {\rm e}^{- E/T} =
\left( {d\tau \over d{\bf p}' \,
dE' } \right)^{-1} \,
{d{\sl n} \over  d{\bf p}' \, dE' }
\, {\rm e}^{- E'/T} \, .
\label{tau/n}
\end{equation}
For simplification, we~next
assume that~2 is much heavier than~1, whereupon we can make no
distinction between relative states of~1 and~2 and
single-particle states~of~1.
Equation~(\ref{tau/n}) then may be also written~as
\begin{equation}
\left( {d\tau \over d{\bf p}_1 \,
dE_{1} } \right)^{-1} \,
{d{\sl n}_{1} \over  d{\bf p}_1 \, dE_{1} }
\, {\rm e}^{- E_{1}/T} =
\left( {d\tau \over d{\bf p}_1' \,
dE_{1}' } \right)^{-1} \,
{d{\sl n}_{1} \over  d{\bf p}_1' \,  dE_{1}' }
\, {\rm e}^{- E_{1}'/T} \, .
\label{ta/n}
\end{equation}

We are now set to address the~scattering of a~particle~3,
representing remainder of the system, with particle~1,
in~the presence of~2.
Transition or
scattering rate
within the system~is, generally, given by a~transition matrix
element squared multiplied by the density of final states in
energy, times~$2 \pi$.  When sampling is carried out over
a~long time, then the
number of transitions within the system from one set of states
to some other should be the same as from the other set to the
first. On~multiplying both sides of~Eq.~(\ref{ta/n}) by
common factors and after manipulations, we~can
demonstrate an~equivalence of the~ergodic
condition~(\ref{ta/n}) with the general condition
of the equality of the number of transitions in the different
directions.  Specifically,
we~multiply both sides of~(\ref{ta/n}) by a~product of average
number
densities~of~3, $dN_3/d{\bf p}_3 \, dE_3 = d{\sl n}_3/d{\bf
p}_3 \, dE_3
\times {\rm e}^{-E_3/T}$, in~the vicinity of two different
states $({\bf p}_3, E_3)$ and $({\bf p}_3' , E_3 ')$, such that
${\bf p}_1 + {\bf p}_3 = {\bf p}_1' + {\bf
p}_3' $ and $E_1 + E_3 = E_1' + E_3'$, and~by a~matrix element
squared in the momentum representation, same for direct and
inverse transitions, $|{\cal M}_{13
\rightarrow 1'3'}|^2 = |{\cal M}_{1'3' \rightarrow 13}|^2
\equiv |{\cal M}|^2$.  After~manipulations we get
\begin{eqnarray}
\nonumber
{d\tau \over d{\bf p}_1 \, dE_{1} }  \, {dN_3 \over d{\bf p}_3
\, dE_{3} } \,  { (2 \pi)^4  \over V^3}   \, & |{\cal M}|^2 &
\, {d{\sl n}_1 \over d{\bf p}_1' \, dE_1'} \,
{d{\sl n}_3 \over d{\bf p}_3' \, dE_3'} \\[.1in]
& = &
{d\tau \over d{\bf p}_1' \, dE_{1}' }  {dN_3 \over d{\bf
p}_3' \, dE_{3}' } \, {(2 \pi)^4 \over V^3} \, |{\cal M}|^2 \,
{d{\sl n}_1 \over d{\bf p}_1 \, dE_1} \,
{d{\sl n}_3 \over d{\bf p}_3 \, dE_3} \, .
\label{Trans}
\end{eqnarray}
The volume~$V$ is assumed here to be large compared to
the range of interactions.
The~l.h.s.~of~(\ref{Trans}) represents number of
transitions that take place, per~element of
energy-momentum
space~$d{\bf p}_1 \, dE_1 \, d{\bf p}_3 \, dE_3 \, d{\bf p}_1'
\, dE_1'$, from states~$13$ to~$1'3'$.  The~r.h.s.~represents
number of inverse transitions per energy-momentum
element~$d{\bf
p}_1' \, dE_1' \, d{\bf p}_3' \, dE_3' \, d{\bf p}_1
\, dE_1$.  By~virtue of energy-momentum conservation the
energy-momentum elements involving any three out of the four
states in the above are
actually identical.  The~number of transitions
in~(\ref{Trans}), e.g.~on the l.h.s.,
is represented in terms of the time spent
by particle~1 in~the region~$({\bf p}_1, E_1)$ times the
probability
of~finding particle~3 in~the region~$({\bf p}_3, E_3)$, times
the transition rate into $({\bf p}_1', E_1')$ and~$({\bf p}_3',
E_3')$.

Equation (\ref{Trans}) is fully equivalent to~(\ref{ta/n})
or~(\ref{dtaun}).  Given that interaction of particle~3 with
any other
particles is perturbative, the~perturbative scattering of~1
and~3 is modified in the presence of~2, compared to free space,
by~a~changed final density of states.  From~(\ref{Trans}) it
follows that, to comply with ergodicity, it~is necessary to
allow particles to participate in~transitions at any time when
in a~given relative state.  In~the transition rate
and, in~consequence, in the cross section, it~is necessary to
account for the changed final-state density.

Within many-body theory, a~final-state density for
scattering
is typically described in terms of
single-particle spectral functions~\cite{kad62,dan84} equal,
up to a~factor, to~the imaginary part of single-particle
Green's function~(\ref{gf}), $A = -2 \, {\rm
Im} \, g$.  For~example, for~the case above, the~number of
transitions in which particles from $({\bf p}_1, E_1)$
and $({\bf p}_3, E_3)$ interact and populate~$({\bf p}_1' ,
E_1')$ and $({\bf p}_3 ', E_3')$, per unit time and per
momentum
volume~$d{\bf p}_3 \, d{\bf p}_1'$, given well-defined
energies of~3, would be represented as
\begin{equation}
{ V \over (2 \pi)^{6}} \, f({\bf p}_3) \, |{\cal M}|^2 \,
A({\bf p}_1', E_1') .
\end{equation}
On~comparing the previous expression for transitions with the
one in terms of~$A$, one can conclude that, in the discussed
case, \begin{equation}
{ V \over (2 \pi)^4} \, A(p_1, E_1) =
 {d{\sl n}_1 \over
d {\bf p}_1 \, dE_1 } =
 {d{\sl n} \over
d {\bf p}_1 \, dE_1 }.
\end{equation}
This implies an integral relation between the density of
relative states~$\rho$ and~$A$,
\begin{equation}
{d{\sl n} \over dE_1} \equiv \rho (E_1) =
{ V \over (2 \pi)^4} \, \int d{\bf p}_1 \, A(p_1, E_1) .
\label{rhoA}
\end{equation}

We now proceed to examine an~explicit form of~$A$
in the ${\cal T}$-matrix approximation in~which interactions
within a~two-particle system, such as that of~1 and~2 in the
above,
are~fully accounted for.  With an~explicit form of~$A$,
we~shall further examine
the validity of~(\ref{rhoA}), establishing a~correspondence
between different terms in~$A$ and in~$\rho$.  We~shall then
discuss in-medium scattering rates.

{}From~(\ref{gf}) we find
\begin{equation}
A({\bf p}_1 , E_1) = - 2
 \, {\rm Im} \,  g  ({\bf p}_1 , E_1) =
{\gamma ({\bf p}_1 ,E_1) \over (E_1 - e(p_1) - u({\bf
p}_1,E_1))^2 + (\gamma({\bf p}_1,E_1))^2/4 } \, .
\end{equation}
Here the real and imaginary parts of self-energy separately
depend on energy and momentum.  In~the ${\cal
T}$-matrix approximation, this dependence corresponds to
a~separate dependence of the ${\cal T}$-matrix on energy and
momentum as in the Lippman-Schwinger equation.
On~expanding~$A$ for low scattering rate~$\gamma$,
in~the discussed case rate for scattering of~1 off~2 given
large~$V$, we~obtain
\begin{mathletters}
\label{A}
\begin{eqnarray}
\nonumber
A({\bf p}_1 , E_1) & \approx &  \left(1 - \left. {\partial
u({\bf p}_1,E_1)
\over \partial E_1} \right|_{E_1={\cal E} ({\bf p}_1)}
\right)^{-1} \, 2 \pi \,  \delta(E_1 - {\cal E} ({\bf p}_1))
\\[.1in]
\label{Aa}
& & -  {\gamma ({\bf p}_1,E_1)} \, {{\cal
P}' \over E_1 - {\cal E}({\bf p}_1)} \\[.1in]
\nonumber
& \approx &  2 \pi \, \delta(E_1 -
{\cal E} ({\bf p}_1))
 +  \left. {\partial u({\bf p}_1, E_1) \over \partial
E_1} \right|_{E_1={\cal E} ({\bf p}_1)}  \, 2 \pi \,
\delta(E_1 - {\cal E} ({\bf p}_1))  \\[.1in] \label{Ab} & &
-
 {\gamma ({\bf p}_1,E_1)} \, {{\cal
P}' \over E_1 - {\cal E}({\bf p}_1)} \, .
 \end{eqnarray}
\end{mathletters}
The energy ${\cal E} ({\bf p}_1)$ in the above is a~solution of
the equation $E _1 - e({\bf p}_1) - u({\bf p}_1,E_1)=0$.
The~factor multiplying the $2\pi \,\delta$ in~(\ref{Aa}) is
termed
a~wave-function renormalization factor, and
\begin{eqnarray}
{{\cal P}' \over x} = {d \over dx} \, {{\cal P} \over
x} =  {d \over dx} \, \lim_{\epsilon \rightarrow
0} \, {1 \over 2} \left( {1 \over x + i \epsilon} + {1 \over x
- i \epsilon}
\right) =  \lim_{\epsilon \rightarrow 0} \, {d
\over
d \epsilon} \, {1 \over 2i}  \left( {1 \over x + i \epsilon} -
{1 \over x - i \epsilon} \right) \, .
\end{eqnarray}
see also~\cite{zim85,sch90}.  Equation~(\ref{Ab}) follows
from~(\ref{Aa}) on recognizing that, according to a~dispersion
relation, the~energy derivative of the mean field is
proportional to the off-shell scattering rate.  Thus, it~is
expected small when the rate is small.  If~$V$ is large enough,
then the mean field itself is small, and~we can further
expand~$A$ in terms~of~$u$,
\begin{mathletters}
\label{AA}
\begin{eqnarray}
\nonumber
A({\bf p}_1 , E_1) & \approx &  2 \pi \,
 \delta(E_1 - e(p_1)) - u({\bf p}_1,e(p_1)) \, 2\pi \,
\delta ' (E_1 -
e(p_1))   \\[.1in] & & +  \left. {\partial u({\bf
p}_1,E_1) \over
\partial E_1} \right|_{E_1=e(p_1)}  \, 2\pi \,  \delta(E_1 -
e(p_1)) -  {\gamma ({\bf p}_1,E_1)} \, {{\cal
P}' \over E_1 - e(p_1)} ,
\label{AAa} \\[.1in] \nonumber
& = & 2\pi \, \left( \delta(E_1 - e(p_1)) - u({\bf p}_1,E_1) \,
\delta ' (E_1 -
e(p_1)) \right) \\[.1in] & & -
 {\gamma ({\bf p}_1,E_1)} \, {{\cal
P}' \over E_1 - e(p_1)} \, .
\label{AAb}
 \end{eqnarray}
\end{mathletters}
Within the ${\cal T}$-matrix approximation, the~mean field and
scattering rate due to~2 are
\begin{eqnarray}
u({\bf p}_1, E_1) = \int {d{\bf p}_2 \over (2\pi)^3} \,
f({\bf p}_2) \, \mbox{Re} \, \langle  p | {\cal T} (E) | p
\rangle  =  \mbox{Re} \, \langle  p_1 | {\cal T} (E_1) | p_1
\rangle / V ,
\label{uu}
\end{eqnarray}
and
\begin{eqnarray}
\gamma ({\bf p}_1, E_1) = -2 \int {d{\bf p}_2 \over (2\pi)^3}
\, f({\bf p}_2) \, \mbox{Im} \, \langle  p | {\cal T} (E)
|  p \rangle = -2 \,
\mbox{Im} \, \langle  p_1 | {\cal T} (E_1)
|  p_1 \rangle /V ,
\label{gam}
\end{eqnarray}
with center expressions following under assumption of
well-defined energies~of~2, and~r.h.s.~expressions representing
results for the specific case under discussion.

On~inserting~$A$ in the form~(\ref{AAa}) into the
r.h.s.~of~(\ref{rhoA}), we~find that the leading term
in~(\ref{AAa})
produces~$\rho_0$ from~Eq.~(\ref{rhoe}).  Second term
in~(\ref{AAa}), with a~derivative of the $\delta$-function,
gives a~complete contribution to~$\Delta \rho$ associated with
the~forward delay time,
\begin{equation}
{d \over d E} \left( \rho_0 \, {1 \over V}
\langle p | {\rm Re} \, {\cal T} (E) | p \rangle \right)
= {1 \over \pi} \sum_{\ell} (2 \ell + 1) \cos{2 \delta_{\ell}}
\, {d \delta_{\ell} \over d E},
\end{equation}
compare~Eq.~(\ref{clasum}).  The~derivative
from the $\delta$-function acting on~$\rho_0$ gives
a~contribution associated with~$\Delta \tau_1^{clas}$,
cf.~Eq.~(\ref{t1def}), and~the derivative acting on~$u$ gives
a~contribution associated with~$\Delta \tau_2^{clas}$,
cf.~Eq.~(\ref{t2def}).  Finally, the~remaining wave-function
renormalization and off-shell terms in~(\ref{AAa}) yield {\em
jointly}, on~insertion into the r.h.s.~of~(\ref{rhoA}) and
after lengthy
manipulations, a~scattering contribution to~$\rho$ of the
form~$\rho_0 \, \sigma v
\, \Delta \tau_s/V$, where~$\Delta \tau_s$ is~given
by~(\ref{tauda}), compare~(\ref{delt}) and~(\ref{erg2}),
indeed confirming the equality in~(\ref{rhoA}).
The~manipulations involve, in~particular, expressing the
$\delta$-function and
principal value in~(\ref{AA}) in~terms of the imaginary and
real parts of the~free two-particle Green's function, and
an~extensive use of the~relations between the~imaginary and
real
parts of the~free two-particle Green's function and the ${\cal
T}$-matrix, following from the~Lippman-Schwinger equation,
Eq.~(\ref{Gal}) with $N=1$ in~$G_0$.  Presence of the
momenta~$p_1$
in the last term in~(\ref{AAa}) and in~(\ref{gam}), far from
the
shell defined by $e(p_1) = E_1$, indicate the effects of
near zone in the interaction of~1 and~2.

Generally, in~a~scattering rate a~single-particle
spectral-function~$A$ would be used for each of final-state
particles.  On~the basis of the example above, one~would
conclude that use of such function leads to the population of
states consistent with two-particle density.  A~more general
relation between~$A$ and~$\rho$, than~(\ref{rhoA}), is
\begin{equation}
{V \over 2 \pi}  \, \int  {d{\bf p} } \,
 {\delta \over \delta f({\bf p}_2)} \,
A \left( {\bf p}_1 ,
E +  P^2/2M  - e(p_2) \right) =
\Delta \rho (E) ,
\label{rhoa}
\end{equation}
where integration is carried out over relative momentum at
fixed total momentum~${\bf
P} = {\bf p}_1 + {\bf p}_2$.  Use~of the distribution~$f({\bf
p}_2)$ in~(\ref{uu}) and~(\ref{gam}) allows for various
momentum values~of~2.  As~$f$ is not normalized to yield one
particle within a~given volume, there is a~possibility for~2
being absent from~$V$, giving a~reduction in the relative
weight
of the correction to the density of states due to the
interaction, cf.~(\ref{AA}).

In~practical applications, the~incorporation of the~first
correction term~in~$A$ in~(\ref{AAa}), associated with~the
forward delay time, amounts to correcting single-particle
energies in the~scattering rates by the mean field.
The~incorporation
of the scattering delays in the final states in scattering
rates can
be much more cumbersome given the form of terms in~(\ref{AAa}).
One~possible solution to get consistency with ergodicity is to
multiply rates or cross sections by factors of the~form
\begin{equation}
1 + \int d{\bf p}_2 \, f({\bf p}_2) \, \sigma v \, \Delta
\tau_s ,
\end{equation}
for each of final-state particles.  That~would be analogous to
putting all delays into the forward delay time in the particle
propagation, discussed in~Sec.~\ref{manipulate}.  A~weakness of
an approach where just final-state densities in two-particle
scattering are modified on~account of scattering with other
particles, is~the disregard of correlations that may persist
throughout the interaction process.  Effects of correlations
become even apparent in a~more detailed analysis of
the~discussed simple
example with perturbative scattering.  Thus a~careful reader
might notice that the matrix elements in~(\ref{Trans}) for
transitions should not~be, generally, taken between plane waves
but rather between eigenstates of an internal hamiltonian
of~1~and~2.  Separation of the wave-functions for these states
into incident and scattered portions would, generally, yield
two-particle and three-particle scattering terms in the
transition rate.  Interference of the forward and scattered
waves for~1 and~2 would yield shadowing corrections in the
matrix element for the scattering of~1 and~3.
The~genuine three-particle
scattering term would be characterized by the lack of momentum
conservation within the subsystem of~1 and~3 alone.  Overall,
such effects are, though, beyond the scope of present paper.

\newpage

\begin{figure}
\caption{
Time delays in the case of a~Breit-Wigner resonance as a
function
of energy from the resonance divided by the width.  The solid line
represents the time for the scattered wave $\Delta \tau_s$.
The short-dashed line represents the time $\Delta \tau_s '$
which
is ergodically consistent when the delay for forward wave or mean
field are neglected.  Finally, the long-dashed line shows the
time delay for the forward wave divided by the fraction of the
incoming wave that is scattered, $\Delta \tau_s '-\Delta
\tau_s$ (cf.~Eq.~(\protect\ref{taufr})).
}
\label{restim}
\end{figure}

\begin{figure}
\caption{
Time delays for a~$\pi^+ p$ system as a function of
c.m.~kinetic energy.  The solid line represents the time delay for
the scattered wave averaged over angles and spin directions,
$\Delta \tau_s$.  The dashed line represents the forward time delay
averaged over spin directions and divided by the fraction of
the incoming wave that is, on the average, scattered,
$\Delta \tau_s '-\Delta \tau_s$.
}
\label{deltatimes}
\end{figure}

\begin{figure}
\caption{
Spin-isospin averaged time delay for $NN$ scattering as
a~function
of the cosine of c.m.~scattering angle, at several values of
laboratory kinetic energy indicated in the figure
in MeV.    }
\label{tausNN}
\end{figure}

\begin{figure}
\caption{
Time delays for $NN$ system as a function of laboratory kinetic
energy.  The solid line represents the time for the scattered
wave $\Delta \tau_s$ averaged over the scattering angle and the
spin and isospin directions.  The dashed line represents the time
for the forward wave, averaged over spin and isospin
directions and divided by the fraction of incident wave
scattered, $\Delta \tau_s' - \Delta \tau_s$.
}
\label{tauNN}
\end{figure}

\begin{figure}
\caption{
Second virial coefficient for nuclear matter, multiplied
by~$2^{5/2}$, from Eq.~(\protect\ref{a2}), as a~function of
temperature.
}
\label{virial}
\end{figure}

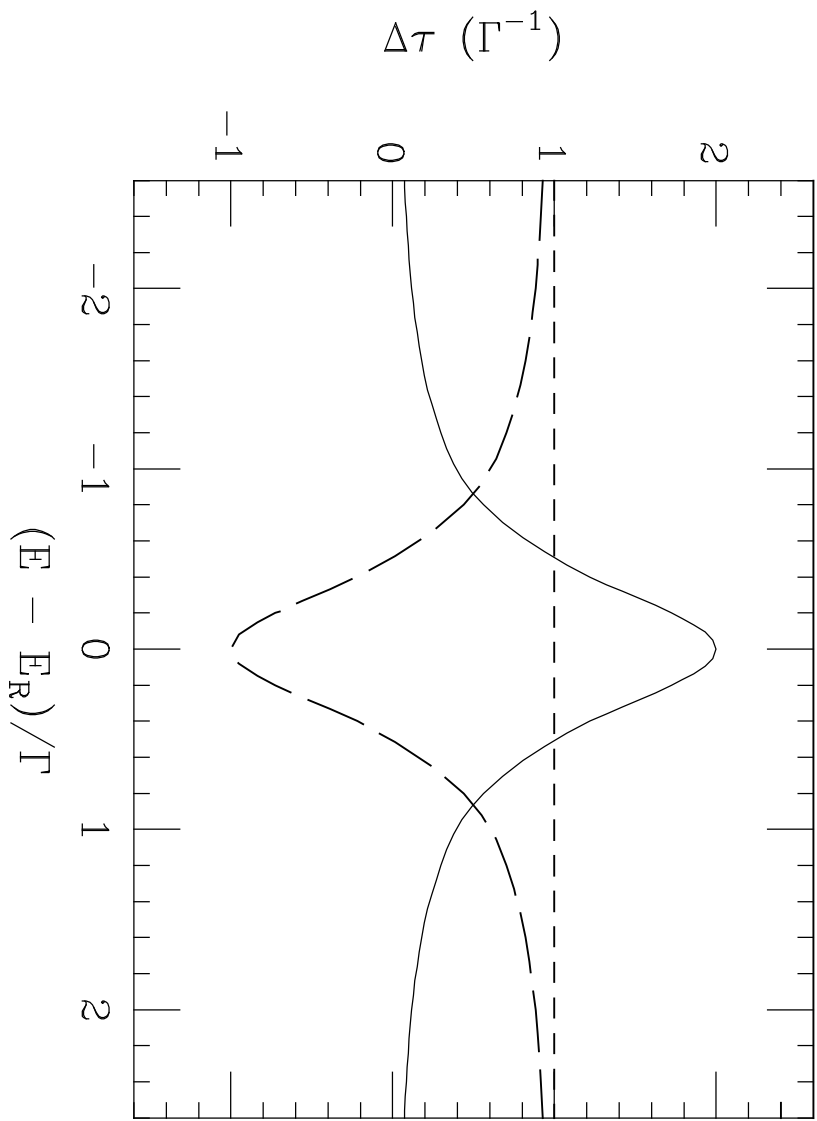
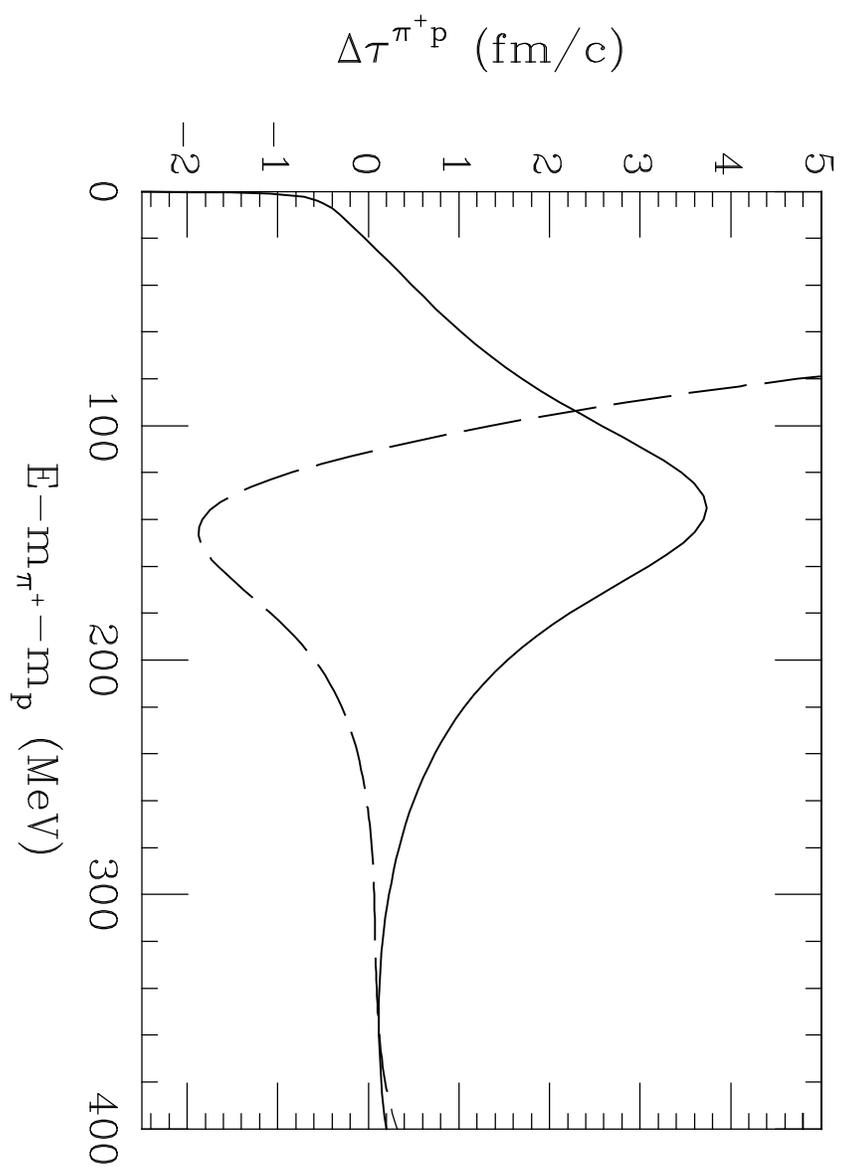
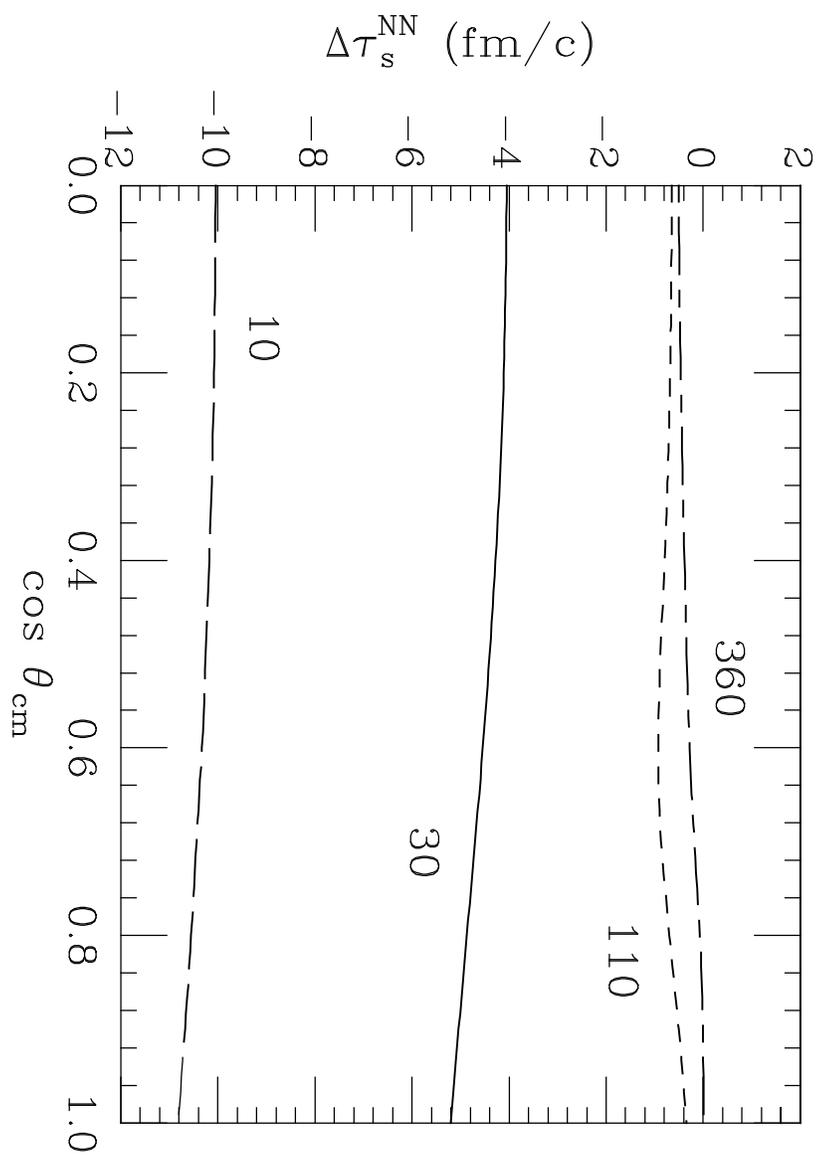
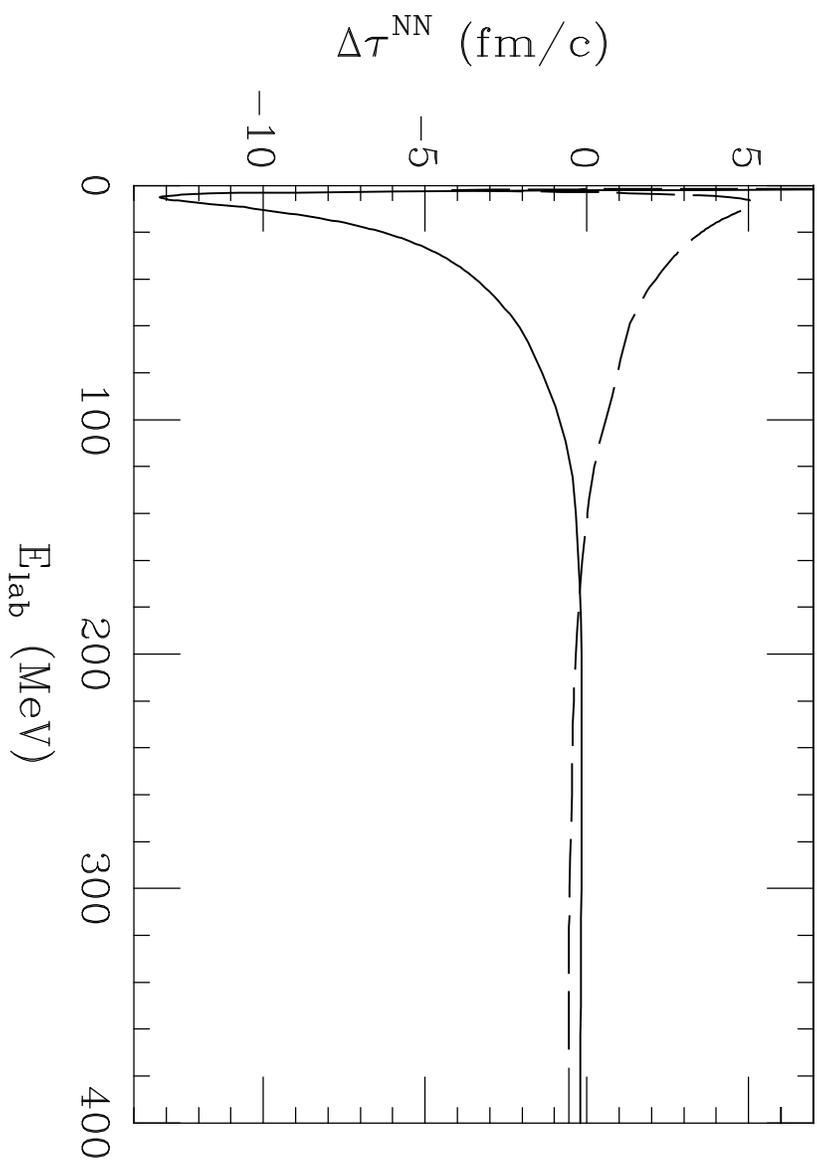
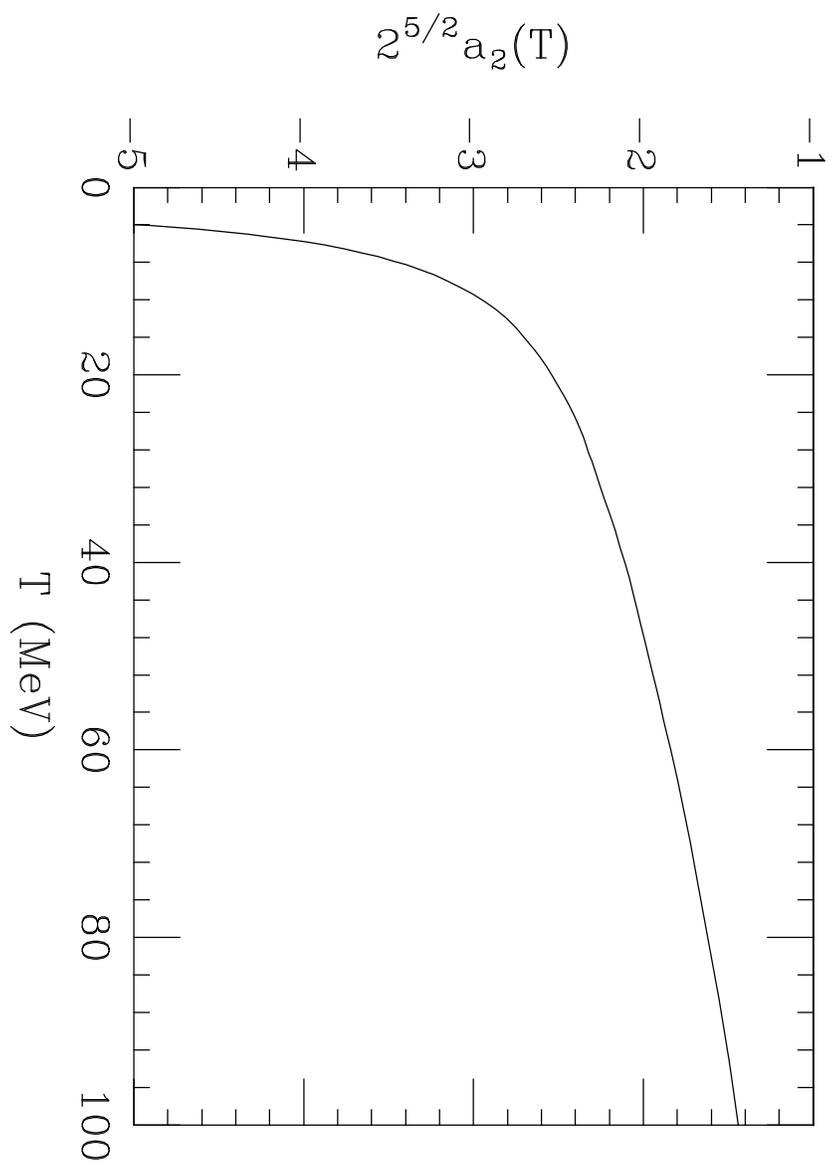
\end{document}